\documentclass[useAMS,usenatbib]{mn2e}

\usepackage{graphicx,natbib,times}
\usepackage{vvo_thesis} 
\usepackage{deluxetable} 
\title[Galaxy Zoo Green Peas]{Galaxy Zoo Green Peas: Discovery of A Class of Compact Extremely Star-Forming Galaxies
\thanks{This publication has been made possible by the participation of more than 200,000 volunteers in the Galaxy Zoo project. Their contributions are individually acknowledged at \texttt{http://www.galaxyzoo.org/Volunteers.aspx}.}}
\author[Cardamone et al.]{
  \parbox[t]{16cm}{
Carolin Cardamone$^{1,2}$\thanks{ccardamone@astro.yale.edu},
Kevin Schawinski$^{2,3}$,
Marc Sarzi$^{4}$,
Steven P. Bamford$^{5}$,
Nicola Bennert$^{6}$,
C. M. Urry$^{2,3}$,
Chris Lintott$^{7}$,
William C.  Keel$^{8}$,
John Parejko$^{9}$,
Robert C. Nichol$^{10}$,
Daniel Thomas$^{10}$,
Dan Andreescu$^{11}$,
Phil Murray$^{12}$,
M. Jordan Raddick$^{13}$,
An\v{z}e Slosar$^{14}$,
Alex Szalay$^{13}$,
Jan VandenBerg$^{13}$
  }\\
$^{1}$Astronomy Department, Yale University 208121, New Haven, CT 06520, U.S.A.\\
$^{2}$Yale Center for Astronomy and Astrophysics, Departments of Physics and Astronomy, Yale University, New Haven, CT 06520, USA \\
$^{3}$Department of Physics, Yale University, P.O. Box 208121, New Haven, CT 06520, USA. \\
$^{4}$Centre for Astrophysics Research, University of Hertfordshire, College ŒLane, Hatfield, Herts AL10 9AB, UK.\\
$^{5}$Centre for Astronomy and Particle Theory, University of Nottingham, University Park, Nottingham, NG7 2RD, UK.\\
$^{6}$Department of Physics, University of California, Santa Barbara, CA 93106, USA. \\
$^{7}$Department of Physics, University of Oxford, Oxford OX1 3RH, UK.\\
$^{8}$Department of Physics and Astronomy, University of Alabama, Tuscaloosa, AL, 35487, USA. \\
$^{9}$ Department of Physics, Drexel University, Philadelphia, PA 19104, USA. \\
$^{10}$Institute of Cosmology \& Gravitation, University of Portsmouth, Portsmouth, PO1 2EG, UK. \\
$^{11}$LinkLab, 4506 Graystone Ave., Bronx, NY 10471, USA. \\
$^{12}$Fingerprint Digital Media, 9 Victoria Close, Newtownards, Co. Down, Northern Ireland, BT23 7GY, UK.\\
$^{13}$Department of Physics and Astronomy, The Johns Hopkins University, Baltimore, MD 21218, USA.\\
$^{14}$Berkeley Center for Cosmological Physics, Lawrence Berkeley National Lab, 1 Cyclotron Road, MS 50-5005, Berkeley, CA 94720, USA\\
}

\newcommand\aj{{AJ}}
\newcommand\araa{{ARA\&A}}
\newcommand\apj{{ApJ}}
\newcommand\apjl{{ApJ}}
\newcommand\apjs{{ApJS}}
\newcommand\aap{{A\&A}}

\newcommand\mnras{{MNRAS}}
\newcommand\pasp{{PASP}}

\pagerange{\pageref{firstpage}--\pageref{lastpage}} \pubyear{2009}

\def\LaTeX{L\kern-.36em\raise.3ex\hbox{a}\kern-.15em
    T\kern-.1667em\lower.7ex\hbox{E}\kern-.125emX}

\begin{document}

\label{firstpage}

\maketitle

\begin{abstract}
We investigate a class of rapidly growing emission line galaxies, known as ``Green Peas,''
first noted by volunteers in the Galaxy Zoo project because of their peculiar bright green colour and small size, unresolved in SDSS imaging.
Their appearance is due to very strong optical emission lines, namely [O III] $\lambda$5007 \AA, with an unusually large equivalent width of up to $\sim$1000 \AA.
We discuss a well-defined sample of 251 colour-selected objects, most of which are strongly star forming, although there are some AGN interlopers including 8 newly discovered Narrow Line Seyfert 1 galaxies.
The star-forming Peas are low mass galaxies (M$\sim10^{8.5}-10^{10} $ ${\rm M_{\odot}}$) with high star formation rates ($\sim 10$ ${\rm M_{\odot}yr^{-1}}$), low metallicities (log[O/H] + 12 $\sim$8.7) and low reddening (${\rm E(B-V) \leq 0.25}$) and they reside in low density environments.
They have some of the highest specific star formation rates (up to $\sim10^{-8}$ ${\rm yr^{-1}}$) seen in the local Universe, yielding doubling times for their stellar mass of hundreds of Myrs.
The few star-forming Peas with HST imaging appear to have several clumps of bright star-forming regions and low surface density features that may indicate recent or ongoing mergers. 
The Peas are similar in size, mass, luminosity and metallicity to Luminous Blue Compact Galaxies.
They are also similar to high redshift UV-luminous galaxies, e.g., Lyman-break galaxies and Lyman-$\alpha$ emitters, and therefore provide a local laboratory with which to study the extreme star formation processes that occur in high-redshift galaxies.
Studying starbursting galaxies as a function of redshift is essential to understanding the build up of stellar mass in the Universe.
\end{abstract}


\section{Introduction}

In this paper, we report on the discovery of an intriguing class of objects
discovered by the Galaxy Zoo project. 
The Galaxy Zoo project \citep{lintottetal2008} has enlisted the help of over 200,000
members of the public to morphologically classify almost $10^6$ galaxies from the
Sloan Digital Sky Survey (SDSS; \citealt{yorketal2000}). The Galaxy Zoo 
website\footnote{\texttt{http://www.galaxyzoo.org}} provides a
randomly selected $gri$ composite colour image from the SDSS main galaxy sample
and asks the volunteers to classify the morphology of the displayed object. One advantage of this distributed approach to
classification is the fact that each object will receive multiple, independent
classifications, and so one can treat the distribution of classifications for
each object in a statistical sense.
These classifications have led to numerous results in galaxy formation and cosmology \citep[e.g., ][]{skibbaetal2008,landetal2008,bamfordetal2009,lintottetal2009,schawinskietal2009,slosaretal2009,dargetal2009a}.

\begin{figure*} 
\begin{center}
\includegraphics[angle=0, width=0.24\textwidth]{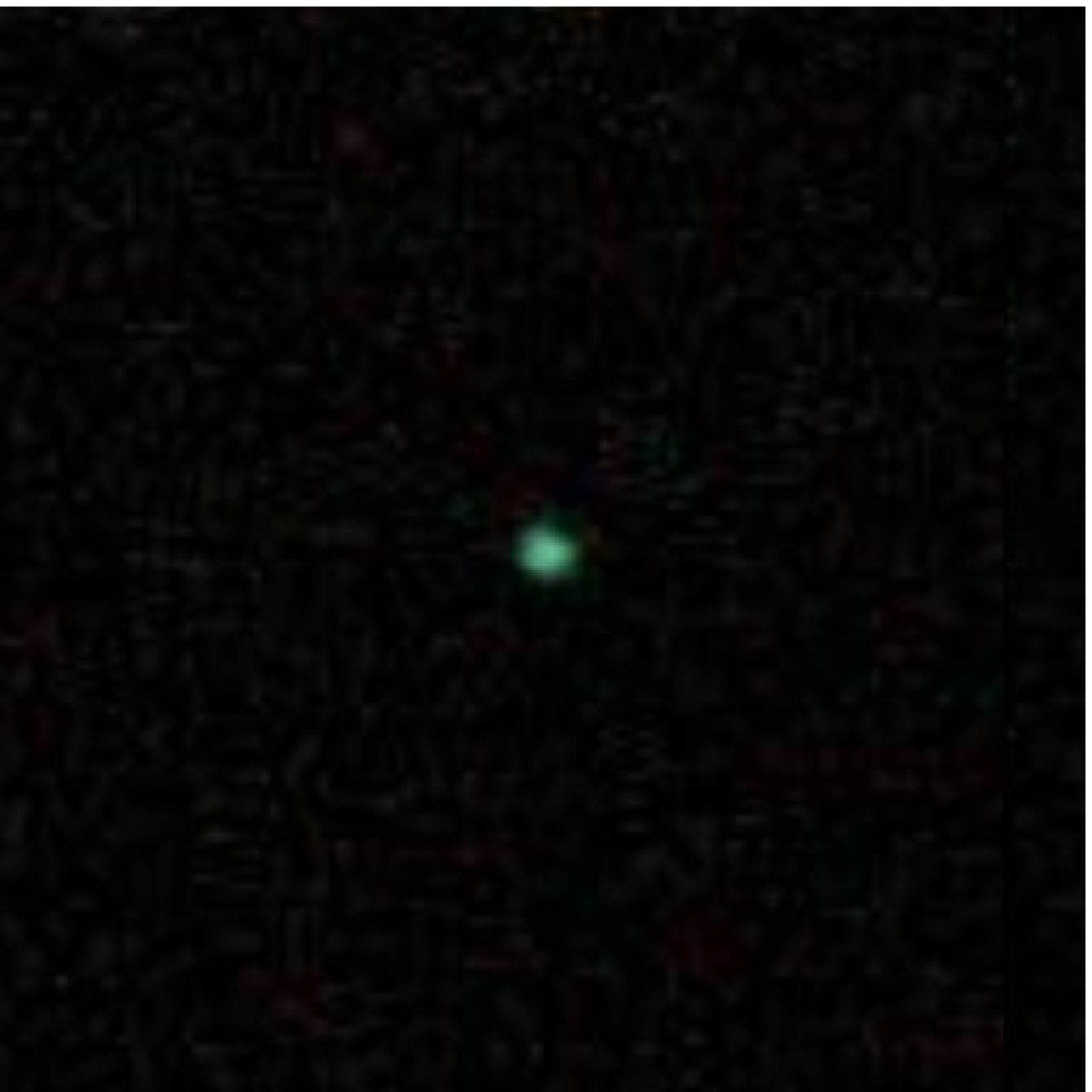}
\includegraphics[angle=0, width=0.24\textwidth]{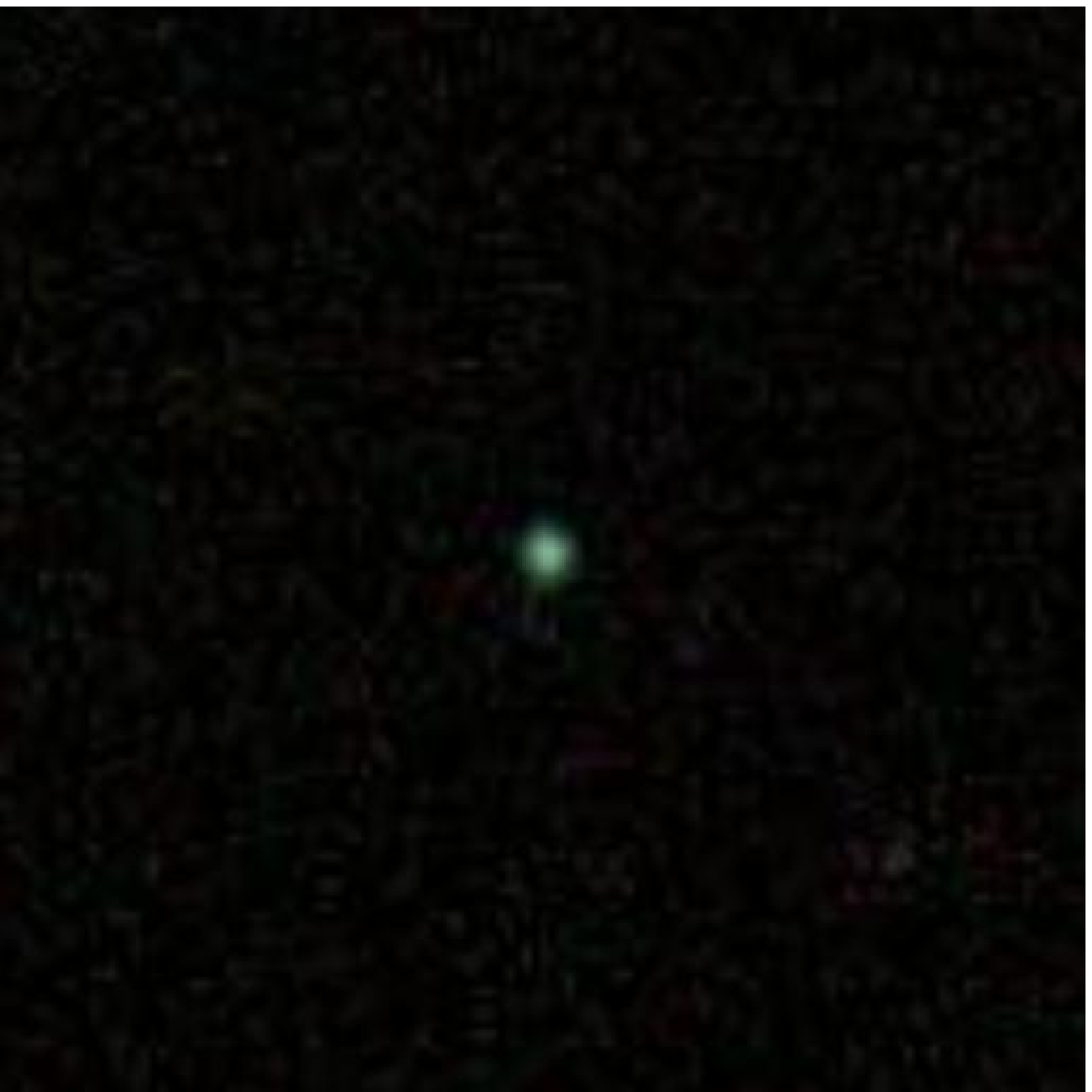}
\includegraphics[angle=0, width=0.24\textwidth]{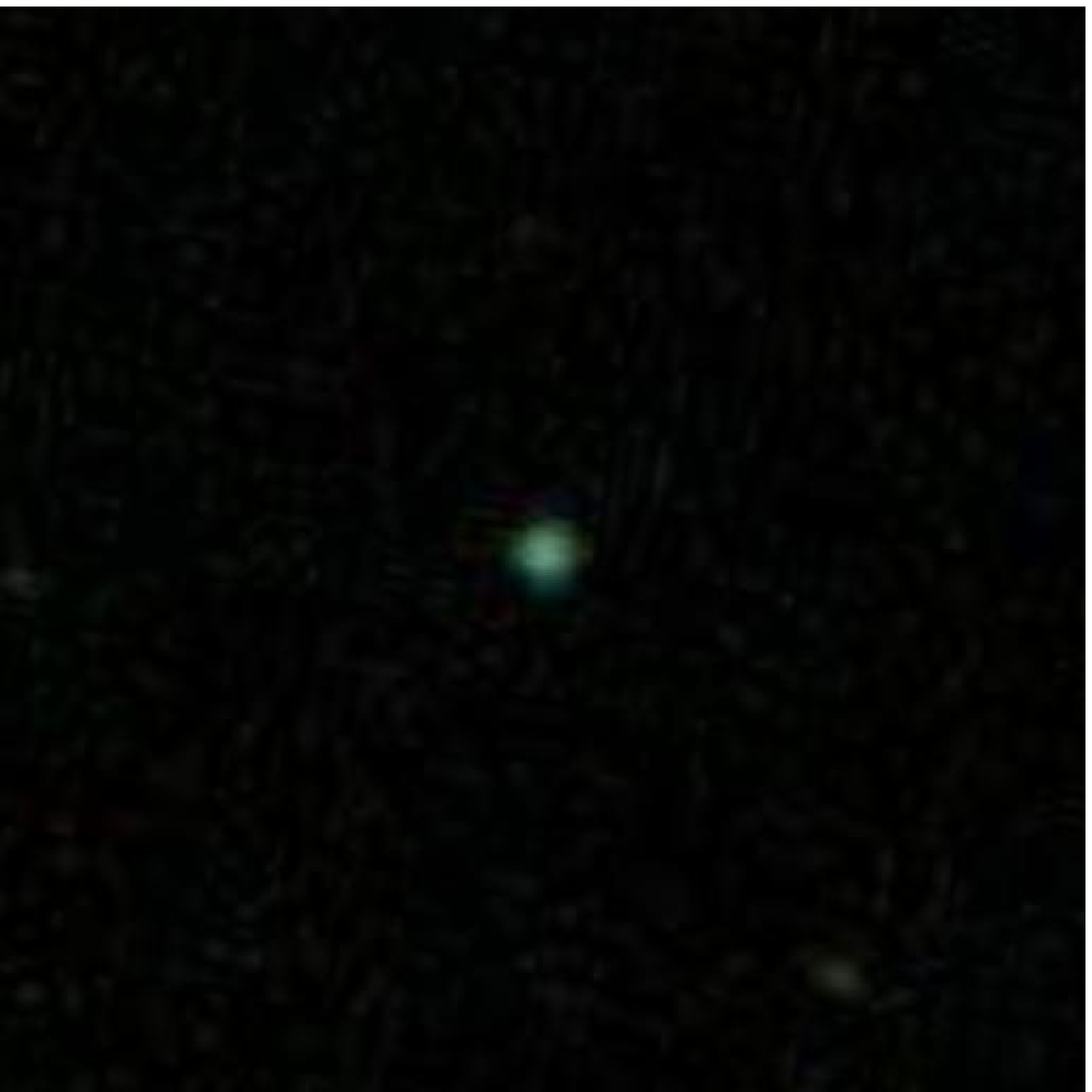}
\includegraphics[angle=0, width=0.24\textwidth]{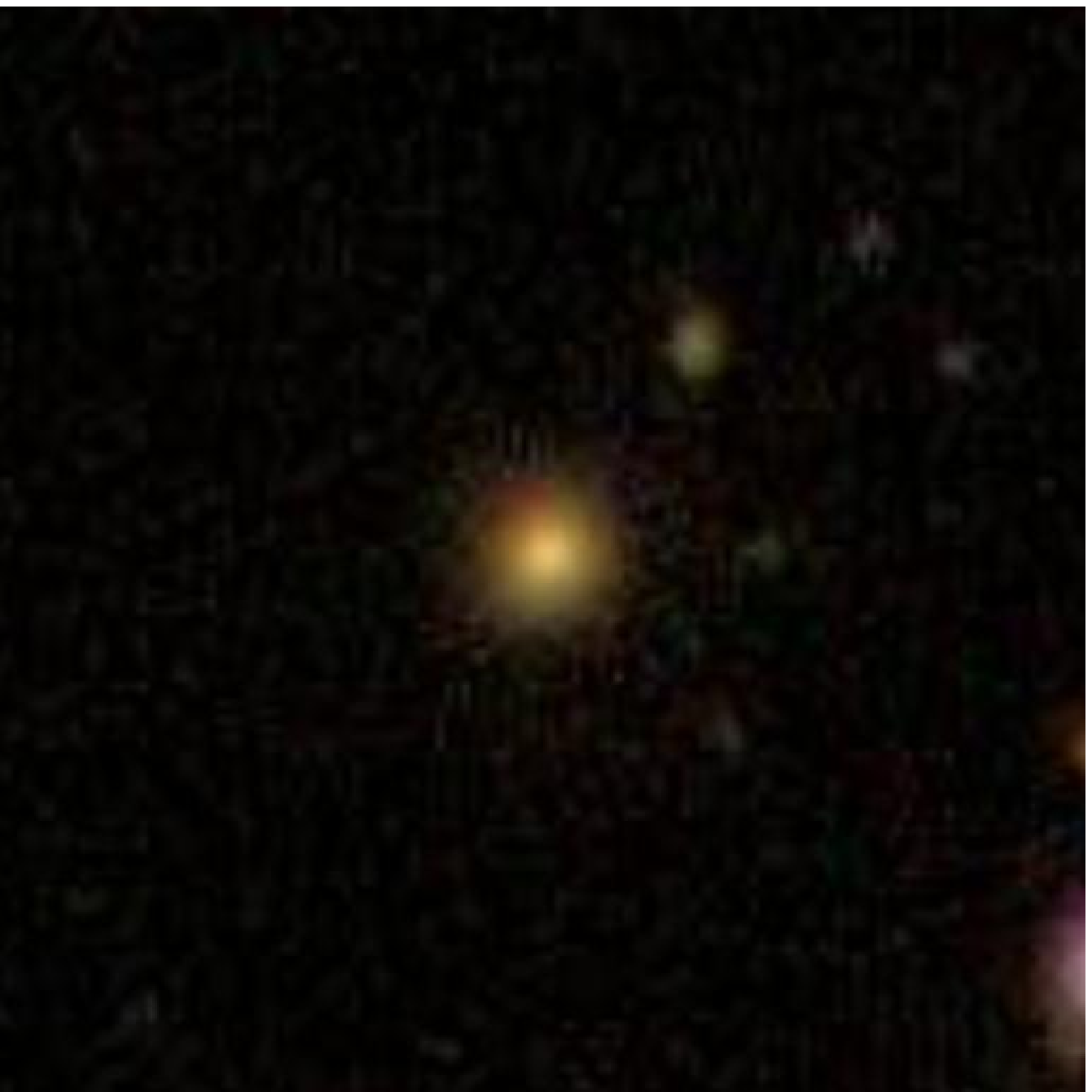}
\caption{Example $g$,$r$,$i$ composite colour 50\arcsec x 50\arcsec~ SDSS images classified by Pea hunters, with the $r$-band representing green light.  The distinctly green colour and compact morphology makes the Peas (left 3 images) easily distinguishable from the classical elliptical (right image).  The elliptical galaxy is clearly red and has a smooth profile, while the Peas are $r$-band dominated and unresolved in these images, appearing like stellar point sources.   All objects shown here are at z$\sim$0.2.}
\label{Peapict}
\end{center}
\end{figure*}

In addition to the website used for classification, Galaxy Zoo also provides an
online discussion forum\footnote{\texttt{http://www.galaxyzooforum.org}} where
volunteers may ask questions about unusual or challenging objects. 
This allows us to tap into another advantage of human classifiers: they can easily identify and then investigate odd objects.  
One such class of highly
unusual objects was named `Green Peas' as they appeared to be
unresolved round point sources that looked green in the $gri$ composite.
Figure \ref{Peapict} shows three example Pea images found by users, as well as a more typical galaxy at the same redshift (z$\sim$0.2).
The volunteers rapidly assembled over a hundred of these objects in a dedicated
discussion thread\footnote{We wish to thank the ``Peas Corps'' for ``giving Peas a chance:'' including, Elisabeth Baeten, Gemma Coughlin, Dan Goldstein, Brian Legg, Mark McCallum, Christian Manteuffel, Richard Nowell, Richard Proctor, Alice Sheppard, Hanny van Arkel.}.
Most of these were classified as stars in the SDSS photometric pipeline \citep{luptonetal2001}.
It quickly became apparent that these objects represent a distinct group: they
all had galaxy-type spectral features (as opposed to broad-line quasar spectra or stellar spectra), and their
green colour was driven by a very powerful [OIII] $\lambda$5007 \AA emission
line that substantially increased the $r$-band luminosity relative to the
adjacent $g$ and $i$ band (green being the colour represented by the
$r$-band in the SDSS colour composites). As a result of this selection, the
Peas are found at redshifts $0.112 \la z \la 0.360$, mostly beyond the main
galaxy sample but much nearer than luminous quasars. 
This discovery
prompted our investigation into the nature of these small [O III]-emitters.

In Section \ref{sec:data}, we present the sample selection, and analyse  their
photometric properties (\S~\ref{sec:sample}), space density (\S~\ref{sec:number}), morphologies in SDSS (\S~\ref{sec:morph}), environments (\S~\ref{sec:env}), spectral properties (\S \ref{sec:spec}-\S  \ref{sec:bpt}) and HST imaging morphologies (\S~\ref{sec:hst}).
The Peas turn out to be largely star-forming objects with some AGN interlopers.
We look into the nature of the few AGN Peas in Section \ref{sec:nls1}.
We study the nature of the star-forming galaxies in Section 
\ref{sec:sf} and
compare them to other known samples of galaxies  in Section \ref{sec:comp}.
Throughout this paper we assume $H_0 = 71  \: {\rm km \: s^{-1} \: Mpc^{-1}}$, $\Omega_{\rm m} = 0.3$ and $\Omega_{\rm \Lambda} = 0.7$, consistent with the \textit{Wilkinson Microwave Anisotropy Probe} 3-year results in
combination with other cosmology probes \citep{spergeletal2007}. 

 \section{Data}
 \label{sec:data}
 \subsection{Sample Selection}
 \label{sec:sample}
Our sample of Peas is taken from the SDSS DR7 spectroscopic sample \citep{abazajianetal2009}.
A survey of a quarter of the sky, the SDSS provides photometry of 357 million unique objects in five filters, $u$, $g$, $r$, $i$ and $z$ \citep{fukugitaetal1996} and spectroscopy of many objects.
Using the CasJobs\footnote{\texttt{http://casjobs.sdss.org/CasJobs/}} application provided by SDSS, we uniformly searched the DR7 spectroscopic sample for Peas (originally noticed by eye in Galaxy Zoo) in the redshift range $0.112 < z < 0.360$ where the [O III] $\lambda$5007 \AA line is in the $r$-band filter.

\begin{figure*}
\begin{center}
\includegraphics[angle=0, width=0.48\textwidth]{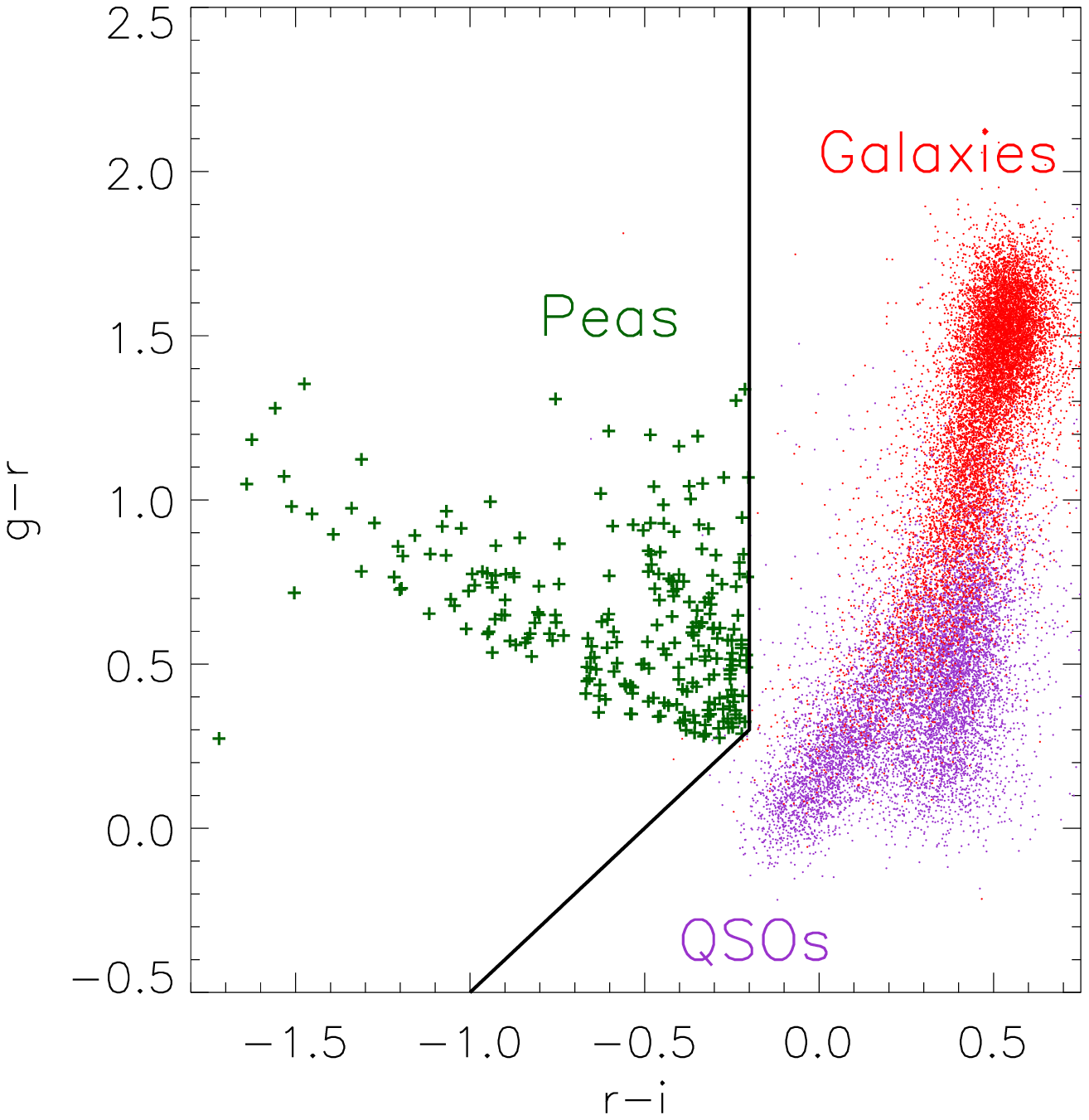}
\includegraphics[angle=0, width=0.48\textwidth]{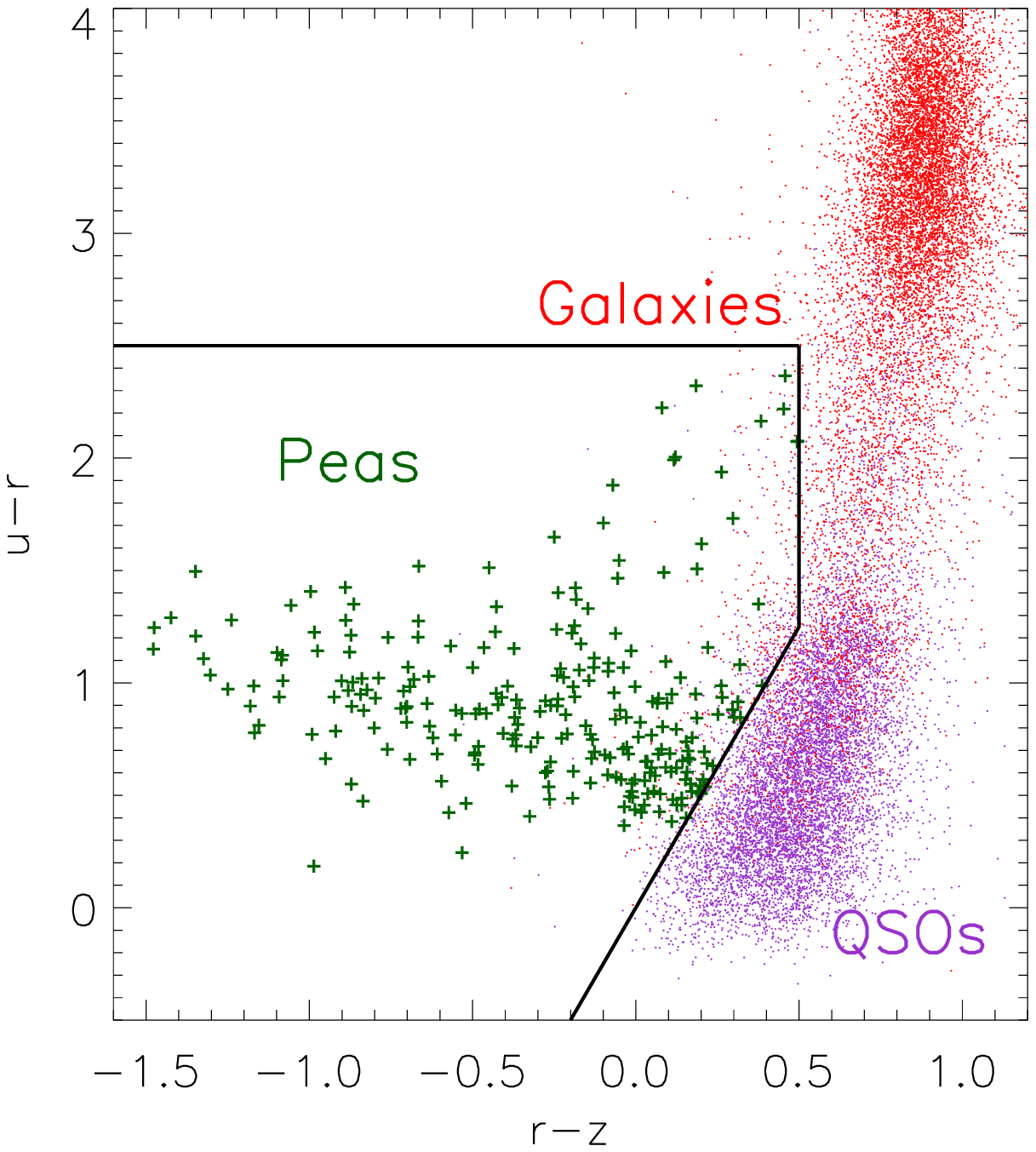}
 \caption{\textit{Left:}  $r-i$ vs $g-r$ colour-colour diagram for 251 Peas (green crosses) and a sample of normal galaxies (red points) matched in $z$ and $g$-band magnitude, and all QSOs (purple points) which lie in the same redshift range as the Peas, $0.112 < z < 0.36$.  
 \textit{Right:}  $r-z$ vs $u-r$ colour for the same classes.  The Peas are most cleanly separated in the  $r-i$, $g-r$ colour diagram, where they stand out as particularly bright in the $r$-band.  The colour cuts were selected to include the Peas identified by the Galaxy Zoo volunteers and to exclude both galaxies and QSOs.}
\label{colmag}
\end{center}
\end{figure*}

To define colour selection criteria, we compared the sample of $\sim$100 Peas identified by the Galaxy Zoo volunteers to a comparison sample of 10,000 galaxies and 9,500 QSOs at the same redshifts over the colour space defined by the 5 SDSS bands.
The 10,000 galaxies were selected to match the redshift and $g$-band-magnitude distributions of the Peas.
The QSO sample contains all spectroscopically confirmed QSOs in the Peas' redshift range, because the QSOs are overall too luminous to match the Peas magnitude distribution.
Figure \ref{colmag} displays two colour-colour plots with Peas (green crosses), comparison galaxies (red points) and comparison QSOs (purple stars). 
The green Pea colour selection is shown by the darkened black lines.  
The precise colour cuts were selected to avoid both the QSO and overall galaxy sequences and to highlight the unusual objects selected by eye in the Galaxy Zoo forum.
The colour limits are:
\begin{equation}
u-r \leq2.5 \\
\end{equation}
\begin{equation}
r-i \leq  -0.2 \\
\end{equation}
\begin{equation}
r-z \leq 0.5 \\
\end{equation}
\begin{equation}
g-r \ge r-i+0.5 \\
\end{equation}
\begin{equation}
u-r \ge  2.5(r-z) \\
\end{equation}
We illustrate the effectiveness of this colour selection in Figure \ref{colmag}. It
divides the Peas from the loci of both the galaxy and the quasar
populations.
This colour selection technique effectively uses the narrow-band survey technique common in high redshift galaxy searches, only here applying it to the broader SDSS filters.
Because we are using broad filters, we are only sensitive to galaxies with extreme [O III] equivalent widths.
Further, because we are using the $r$-band filter for colour selection, we are not sensitive to Pea-like objects at lower and higher redshifts.
As seen in Figure \ref{colmag}, the Peas do indeed have distinct colours from the SDSS main galaxy sample.
This is especially noticeable in the $r-i$ and $g-r$ colours used to create the 3-colour images.
Using the colour selection, we find a sample of 251 Peas taken from all SDSS spectroscopic galaxies whose spectra we further analyse to understand their properties.
 
Although most of the Peas were identified by the SDSS spectroscopic pipeline as having galaxy spectra (4 were classified as unknown), only 7 were targeted by the SDSS spectral fibre allocation as galaxies.  
Most were targeted as serendipitous objects, with the majority flagged as  
${\texttt{ SERENDIP\_BLUE}}$, $\texttt{SERENDIP\_DISTANT}$ and $\texttt{TARGET\_QSO\_FAINT}$.
These target flags are for objects lying outside the stellar locus in colour space and \texttt{DISTANT} here refers to distance from the stellar locus, $\texttt{QSO\_FAINT}$  is also used for objects flagged as stellar that are both fainter than $i$=20 mag and outside the stellar locus in colour space  \citep{stoughtonetal2002}.
These objects were targeted by fibers as they became available in a given field, so their selection function is not well determined.
Without a uniform selection for the Peas across the sky, their
absolute space density cannot be accurately assessed.  

\subsection{How common are the Peas?}
\label{sec:number}
Because the spectroscopic selection is biased in unquantifiable ways, i.e., based on the availability of a spectral fibre in a given pointing, the space density of the Peas is difficult to assess.
In order to estimate the space density of the Peas in SDSS, we need to search the entire SDSS photometric catalogue for objects with our colour selection criteria.
We note that the Pea colour selection can also return much higher redshift objects by finding other emission lines in the $r$-band filter.
Therefore we first search the spectroscopic sample to help understand these contaminants.
Dropping the Pea redshift selection limits, we find 198 objects at higher redshift in the spectoscopic sample which fall into our colour-selection region.
These are mostly QSOs (only 4 have a spectral type labelled as \texttt{galaxy} by the SDSS pipeline), which cluster at redshifts z=1.2, where the 2800\AA~Mg II line falls into the $r$-band, and z=3.0, where the 1546~\AA~ CIV line falls into the $r$-band.
Very roughly, in the entire spectroscopic database, there are comparable numbers of Peas and higher-z QSOs in the colour-selection region.
Searching the entire SDSS photometric catalogue (\texttt{PhotoObj}) regardless of spectroscopic information, 
 we limit ourselves to objects with the same $r$-band magnitude range as the Peas (18 $\leq r \leq $20.5) 
 and to objects with similar compactness (\texttt{petrorad\_r} $\leq$ 2.0 $\arcsec$), in order to limit the contaminants from other redshifts.
 We further added the requirement of small $g$, $r$ and $i$ band photometric errors to avoid the scattering of contaminants with poor photometry into the colour selection region.
This search returns 40,222 objects.
The unique area of the SDSS DR7 footprint covers 8423 sq. deg, so this implies a rough spatial density estimate of 5 per square degree.
Strictly speaking this is an upper limit because our selection likely still contains QSOs from higher redshifts.
Looking at QSO number counts, we would expect to see $\sim$3 per sq. degree in our magnitude range \citep{richardsetal2005}, leaving two Peas per square degree brighter than 20.5 magnitudes.
Therefore, we conclude the Peas are rare objects.

\subsection{Morphology from SDSS imaging}
\label{sec:morph}
Of the 251 Peas in our sample, 215 were classified as morphological type \texttt{STAR} and not as extended objects by the SDSS pipeline.  
Compared to the size of the galaxies as measured by the SDSS Pipeline (Petrosian radius: \citealt{blantonetal2001, yasudaetal2001}), they are significantly smaller.
This is expected because the spectroscopic galaxy sample was selected to be both resolved and brighter than the faintest Pea \citep{straussetal2002}.  
The typical resolution of the SDSS images is large ($\ga 1 \arcsec$), just below the peak of the sizes of the Peas 
as measured by the SDSS Petrosian radius.
At the typical redshifts of the Peas, this angular scale corresponds
to an upper limit on the physical half-light radius of approximately 5 kpc.

\subsection{Environment}
\label{sec:env}
\begin{figure}
\includegraphics[angle=270,width=0.48\textwidth]{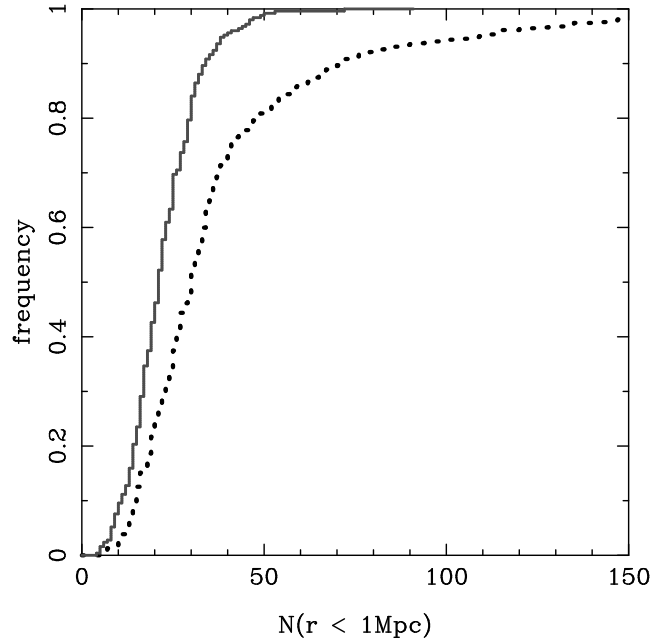}
\caption{The cumulative distribution of neighbour counts for the Peas (solid line) is significantly different than a comparison galaxy sample (dotted line).  No Peas live in the highest density environments.}
\label{env}
\end{figure}

Because the Peas are at redshifts higher than the main spectroscopic galaxy sample, we measure the projected densities around the Peas and a sample of random galaxies matched in both luminosity and redshift, counting the number of projected neighbours within a radius of 1 Mpc and brighter than an absolute $i$-band magnitude of -20.5 at the redshift of the target are counted.
This magnitude limit was selected to be well above the detection limit of the highest redshift Peas.
Figure \ref{env} shows the cumulative distribution of neighbour counts for the Peas (solid line) and the control sample of galaxies (dotted line).
Although this is a simple test, neglecting foreground and background contamination in the neighbour counts, and hence underestimating the difference between high and low density environments, it is clear that the Peas inhabit significantly {\it lower} density regions than typical galaxies of the same $i$-band luminosity.  A Kolmogorov-Smirnov test indicates that the difference is significant at a greater than 8-$\sigma$ level.
We find the median environmental density around the Peas is less than two-thirds of that around normal galaxies.

\subsection{Spectral Analysis}
\label{sec:spec}
We downloaded all of the Peas' spectra from the DR7 archive and carefully re-fit them, paying close attention to both the continuum and emission lines.
We fit the stellar continuum and ionised-gas emission, following the technique of \citet{sarzietal2006}, using the corresponding  \textsc{PPXF}  (\textsc{Penalised Pixel Fitting}; \citealt{cappellariemsellem2004}) and \textsc{GANDALF (Gas AND Absorption Line Fitting)} IDL (Interactive Data Language) codes adapted for dealing with SDSS data\footnote {Both codes can be downloaded from \texttt{http://www.strw.leidenuniv.nl/sauron/}}.
Stellar population templates \citep{tremontietal2004} and Gaussian emission-line templates were simultaneously fitted to the data.
When it improved the overall fit (9 cases), the Gaussian emission line templates included both broad-line and narrow-line components for the Balmer Series.

\begin{figure}
\includegraphics[angle=0,width=0.48\textwidth]{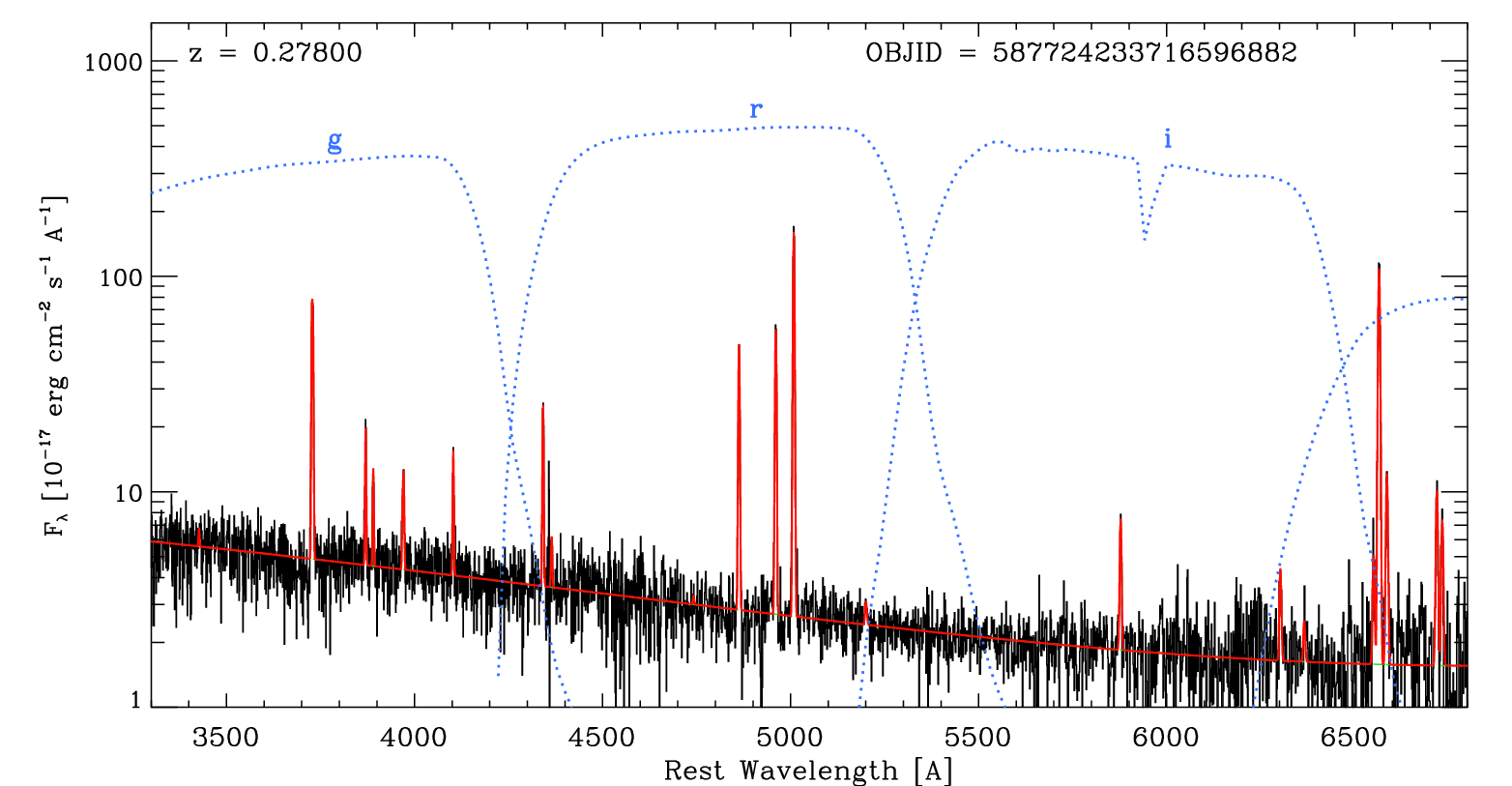}
\includegraphics[angle=0,width=0.48\textwidth]{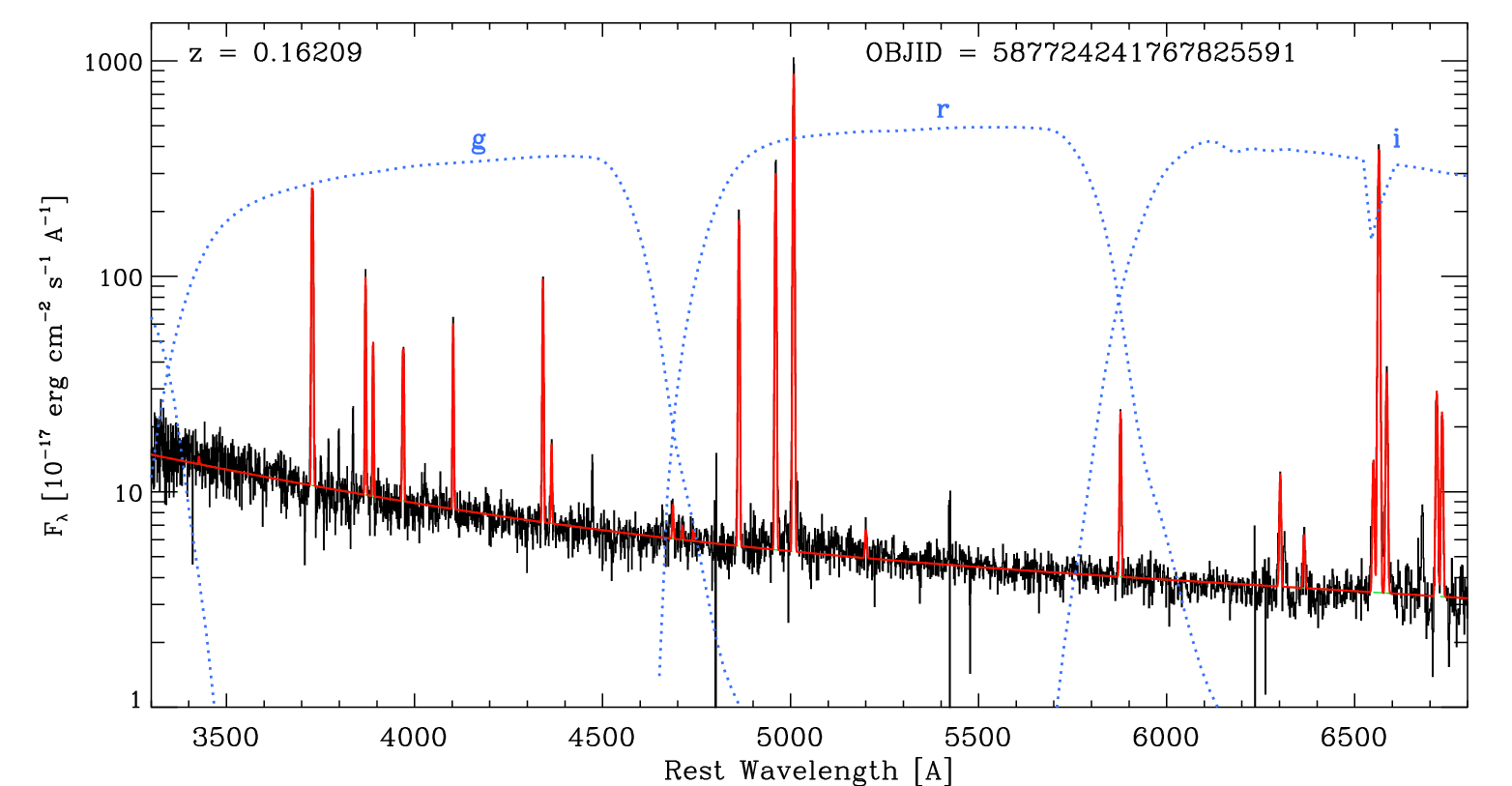}
\includegraphics[angle=0,width=0.48\textwidth]{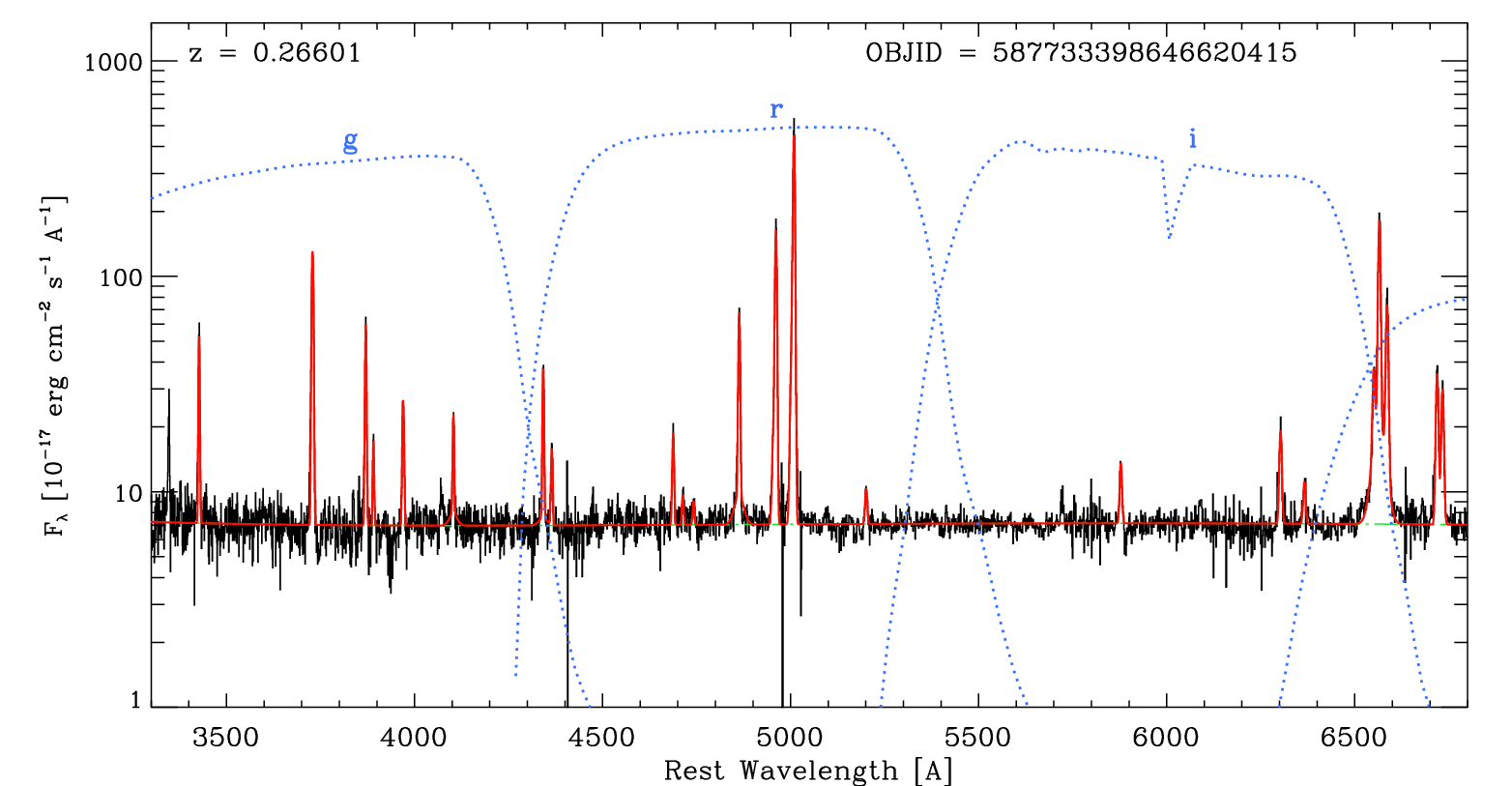}
\caption{Example spectral fits from GANDALF. The top two plots show typical star-forming Peas and the bottom plot shows a typical Narrow Line Seyfert 1. In black is the rest-frame observed spectrum and in red the fit from GANDALF.  The SDSS filter band passes are included as blue dotted line, shifted into the rest-frame of the Pea.  Notice in all examples, the [O III] $\lambda$5007 \AA~ line is redshifted inside the $r$ bandpass. The Peas show very strong emission lines, including clear lines detected throughout the Balmer series.}
\label{fits}
\end{figure}

To ensure acceptable fits, we limit our spectroscopic sample to those Peas with a S/N greater than 3 in the continuum near the H$\beta$ and H$\alpha$ regions (specifically, measuring the S/N in the bands 6350-6500\AA~  and 5100-5250\AA). 
Additionally, we limited our emission-line classification sample to those Peas with a S/N $ \ge 3$ detection in each of the emission lines: H$\alpha$, H$\beta$, [O III] $\lambda$5007 \AA and [N II] $\lambda$6583 \AA, following \citet{kauffmannetal2003}.
We note that for many of our objects near z$\sim$0.3, sky lines fall on top of the [O III] line and the H$\beta$ line, and we removed all of these objects from our sample.
One of these objects has broad Balmer lines, with both sky lines falling inside the H$\beta$ profile.
This object is identified as a Narrow Line Seyfert 1 (NLS1) in the literature \citep{zhouetal2006}, but we do not consider it further in this paper.
These cuts result in a sample of 103 narrow-line objects to be further analysed (see \S~\ref{sec:bpt}) plus 
nine objects best fit by a two-component Gaussian in the Balmer lines.
The width of the broad Gaussian components ranged from just over 600 ${\rm km}$ ${\rm s}^{-1}$ to $\sim$5000 ${\rm km}$ ${\rm s}^{-1}$.
Eight of these objects have FWHM narrower than 2000 ${\rm km}$ ${\rm s}^{-1}$ and are thus classified as Narrow Line Seyfert 1s (see  \S\ref{sec:nls1}).
These eight SDSS spectroscopic objects have not previously been identified as NLS1s, \citep[e.g., ][]{williamsetal2002,zhouetal2006}.
Three examples of spectral fits are shown in Figure \ref{fits}.
The top 2 spectra are of narrow line objects and the bottom spectra shows a NLS1 object.
In all three cases one can easily see a prominent [O III] doublet near $5007$ \AA.

In summary, we have 9 Peas with two component fits, (1 broad Line Seyfert 1 and 8 Narrow Line Seyfert 1s) and 103 narrow line Peas.

\subsection{Spectral Classification}
\label{sec:bpt}
\begin{figure}
\includegraphics[angle=0,width=0.48\textwidth]{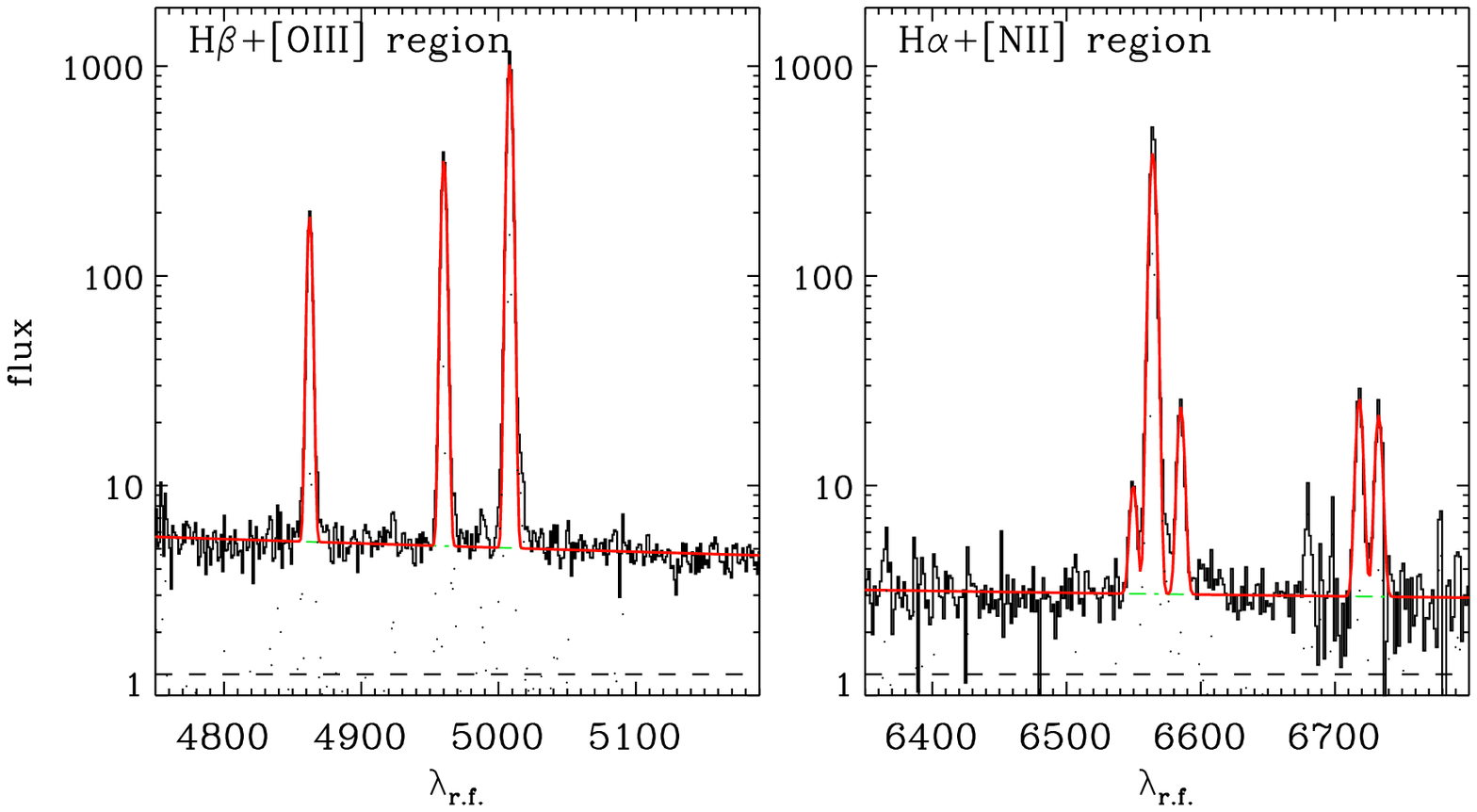}
\includegraphics[angle=0,width=0.48\textwidth]{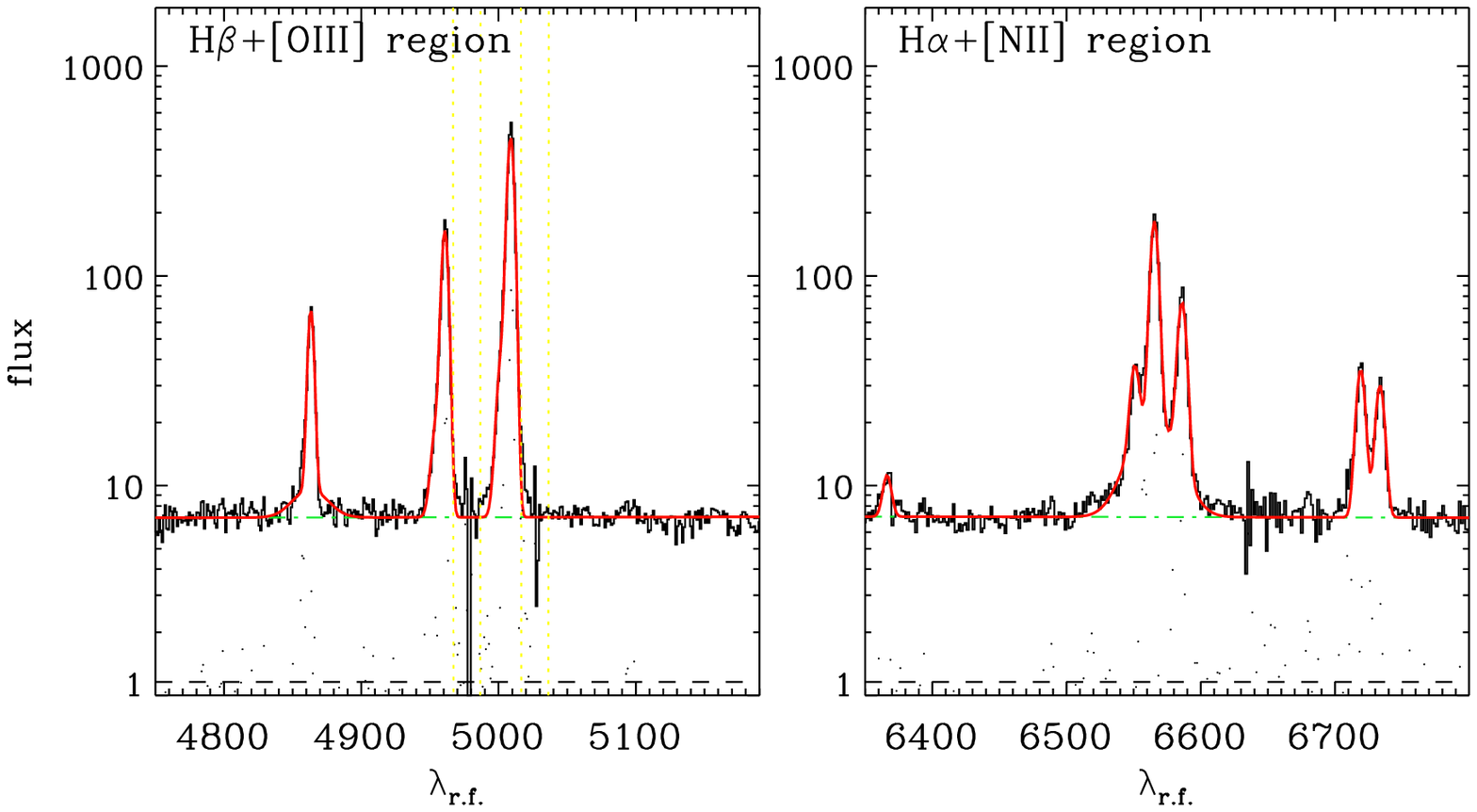}
\caption{Example spectral fits from \textsc{GANDALF} in the regions around H$\alpha$ (right) and H$\beta$ (left).  The top plots show a typical star-forming Pea and the bottom plots show a typical Narrow Line Seyfert 1.}
\label{fitscut}
\end{figure}

The SDSS spectra cover the observed range 3800--9200 \AA~at a resolving power of R$\sim$1800.
At the Peas' redshifts, this range includes the regions around both the H$\beta$ and H$\alpha$ spectral lines.
Example fits to these lines are shown in Figure \ref{fitscut}.

\begin{figure}
\begin{center}
\includegraphics[angle=0, width=0.48\textwidth]{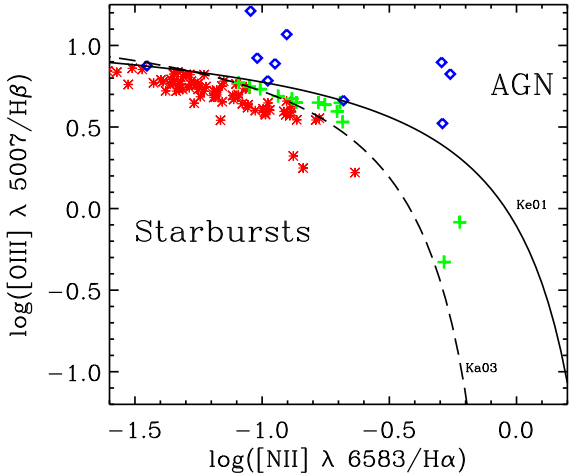}
\caption{The BPT diagram classifies 103 narrow-line Peas (all with S/N $\ge$3 in the emission lines) as 10 AGN (blue diamonds), 13 transition objects (green crosses) and 80 starbursts (red stars). Solid line: \citet{kewleyetal2001} maximal starburst contribution (labelled Ke01); Dashed line: \citet{kauffmannetal2003} line separating purely star-forming objects from AGN (labelled Ka03).  The clustering of the starbursts in the top left corner of the plot indicates that they likely have low metallicity. }
  \label{bpt}
\end{center}
\end{figure}

We use a classical emission line diagnostic originally devised by Baldwin, Phillips
\& Terlevich 1981 (hereafter BPT) and modified by others \citep{osterbrockandpogge1985, veilleuxandosterbrock1987, kewleyetal2001, kewleyetal2006, kauffmannetal2003} to classify the narrow line objects.
Emission line diagnostics probe the nature of the dominant ionising source and separate the galaxies dominated by ongoing star formation from those dominated by non-stellar processes (namely Seyfert and LINER galaxies).
The classical BPT diagram, which compares the ratio of [O III] $\lambda$5007 \AA/H$\beta$ to [N II] $\lambda$ 6583 \AA/H$\alpha$, has been shown to be an efficient measure of the ionising source in a galaxy \citep{kewleyetal2006}.
Additionally,  \citet{kewleyetal2001} calculated the maximum starburst contribution from theoretical models, including modern stellar population synthesis, photoionisation and shock models (labelled Ke01 in Figure \ref{bpt}).
\citet{kauffmannetal2003} empirically shifted this line to separate purely star-forming objects from the rest using a set of $\sim23,000$ SDSS spectra (labelled Ka03 in Figure \ref{bpt}).
Objects that lie in between the Ke01 and Ka03 lines are transition galaxies containing a mix of star formation and a central AGN component (Kewley 2006).
We note that the SDSS spectral fiber includes only the light of the central 3\arcsec~of the galaxy, so emission originating from the central AGN-ionised regions can be mixed with emission originating from extended star formation.

\begin{table}
\caption{Spectroscopic Classification}
\label{tab:classify}
\begin{tabular}{@{}lc}
\hline
 \multicolumn{1}{c}{Type} & \multicolumn{1}{c}{Number} \\
\hline
\hline
Broad Line AGN (Seyfert Type 1) & 1 \\
Narrow Line Seyfert 1s & 8 \\
Narrow Line AGN (Seyfert Type 2) & 10 \\
Transition Objects & 13 \\
Star Forming & 80 \\
\hline
Total & 112
\end{tabular}
\end{table}

Analysis of the narrow-line Peas' spectra (Figure \ref{bpt}) revealed that the majority of the objects are star forming (80, red stars), but there are also 10 Seyfert 2 (blue diamonds) and 13 transition objects (green crosses).
The location of the star-forming galaxies in the top left of the BPT diagram indicates they likely have lower metallicity.
This sample of galaxies is discussed in greater detail in \S~\ref{sec:sf}.  
In Table \ref{tab:classify}, we summarise the spectroscopic identifications.

\subsection{Morphology from HST imaging}
\label{sec:hst}
Because the SDSS resolution is too low to measure the actual sizes of the Peas, we searched the HST archive, MAST\footnote{Multimission Multimission Archive at STScI, \texttt{http://archive.stsci.edu/index.html}}, finding 5 public images. The observations are summarized in Table \ref{tab:hst}.

\begin{figure*}
\includegraphics[angle=0,width=0.68\textwidth]{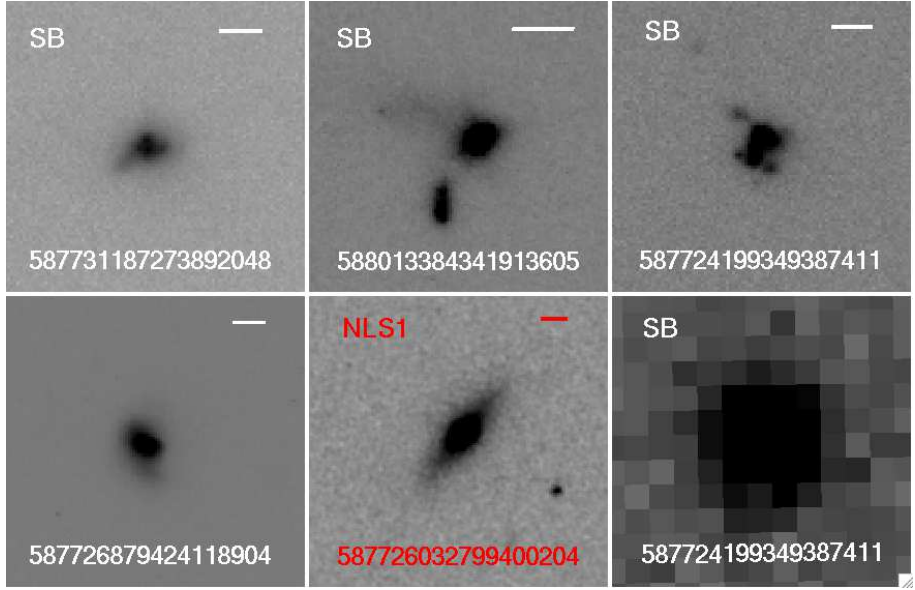}
\caption{\textit{Hubble Space Telescope} images of all Peas available
in the archive. In each panel, we indicate a physical scale of 3 kpc at the
redshift of the object (bar in top right of image). In the top row, we show three Peas classified as
actively star-forming by their emission line ratios. In the bottom row, from
left to right, an unclassified Pea, a Pea classified as a NLS1 and an SDSS
image of a star-forming Pea at the same scale to highlight the degree to which
the Peas are unresolved in typical SDSS imaging. 
These HST images illustrate that the upper
limit on the Peas' physical scale based on ground-based SDSS imaging is a
significant overestimate of the true physical size.  
Furthermore, these HST images reveal that
the star-forming Peas exhibit a complex morphology indicative of significant
disturbances that may be due to mergers and/or clumpy regions of star formation or extinction.}
\label{hst}
\end{figure*}

We reduced the ACS data, starting with the pipeline-reduced *flt* images
and using \textsc{multidrizzle} to remove cosmic rays and defects,
correct for distortion, and improve sampling of the PSF
with a final scale of $0.04\arcsec/{\rm pixel}$.
In the case of the WFPC2 data,
we started with the pipeline-reduced *c0f* images
and used \textsc{drizzle} with parameters pix.scale=0.5 and pix.frac=0.8,
leading to a final scale of 0.023 $\arcsec$/pixel for the PC chip
and 0.05 $\arcsec$/pixel for the WF4 chip, respectively.
Figure \ref{hst} shows all 5 public HST images of the Peas compared to a single typical 
ground-based SDSS image (bottom right).

For the ground-based image (Fig \ref{hst}; bottom right), the SDSS archive lists a Petrosian
radius of 1\farcs8, roughly 8 kpc at the Pea's redshift (z=0.2832).
This object is clearly unresolved when compared to the HST image of the same galaxy (top right) and the size as measured by the SDSS pipeline is therefore an upper limit.
Of the 5 Peas with HST data, one is classified as a NLS1 (see \S \ref{sec:nls1}; bottom centre image), three
are identified as star forming (see \S \ref{sec:sf}; top images) and the last one has sky lines over the [O III] region and is thus unclassified (see \S \ref{sec:spec}; bottom left).
The three star-forming galaxies were imaged as part of a study of local UV-luminous galaxies
\citep{heckmanetal2005,overzieretal2008}.
All three Peas classified as starburst galaxies reveal complex structures too small to be resolved in ground-based imaging.
The morphology of the top right object for example, shows several ``knots" instead of one central component. 
These knots may be different star-forming regions, suggesting a morphology typical of merger events.
Although the Peas live in low density environments, their star formation could still be driven by merging activity.
The top centre image shows a central component with extended structures that look like two tidal
tails reaching out to the east and south, possibly connecting to another galaxy in the south.
The centre of the top left image seems also to consist of at least two knots rather than a smooth overall light distribution.
The NLS1 (bottom centre) is from a study of AGN host galaxies \citep{schmitt2006} and the last
object (bottom left) was imaged serendipitously in a study of Kuiper belt objects \citep{noll2007}.
In contrast to the star-forming galaxies, the NLS1, bottom centre, looks like a spiral seen edge on.
This AGN aside, all of the HST images reveal complex structures much too small to be resolved in
ground based imaging.
Although our statistics are too low to make any conclusions
on the general nature of the Peas, this is an interesting trend.

\begin{table*}
\caption{HST Images}
\label{tab:hst}
\begin{tabular}{@{}lrr l ll ll l}
\hline
\hline
\multicolumn{1}{c}{SDSS Obj ID} &\multicolumn{1}{c}{RA (J2000)} &\multicolumn{1}{c}{Dec(J2000)} &\multicolumn{1}{c}{z} &\multicolumn{1}{c}{Instrument} & \multicolumn{1}{c}{Chip} & \multicolumn{1}{c}{Filter} & \multicolumn{1}{c}{Exposure Time (s)} & \multicolumn{1}{c}{Figure location}\\
\hline
587731187273892048    & 351.41345 &   0.75201  & 0.2770  & WFPC2 &     PC    & F606W  & 3600 & top left \\
588013384341913605    & 141.50168 & 44.46004  & 0.1807  & ACS       &     WF   & F850LP & 2274 & top centre \\
587724199349387411    &   10.22636 & 15.56935  & 0.2832  & WFPC2 &     PC    & F606W  & 3600 & top right \\
587726879424118904    & 344.49532 &  -8.62438  & 0.3081  & ACS       &     WF   &  clear   & 1071 & bottom left \\
587726032799400204    & 211.91701 &   2.29671  & 0.3092  & WFPC2 &     WF4 &  F814W &   1200 & bottom centre \\
\hline
\end{tabular}
\end{table*}

\section{The Properties of Narrow Line Seyfert 1 Peas}
\label{sec:nls1}	
NLS1s constitute  $\sim$15\% of low redshift (z$\leq$0.5) Seyferts \citep{williamsetal2002} and are characterised by H$\beta$ line widths broader than classical narrow-line AGN but narrower than classical Seyfert 1 galaxies.
They appear to have Eddington ratios near 1 and black hole masses below the typical M$_{\rm BH}$-$\sigma$ relation,
suggesting a time delay between the growth of the galaxy and the growth of the central black hole mass \citep{ryanetal2007}.
Several authors use the [OIII] line as a surrogate for the bulge stellar velocity dispersion $\sigma$, assuming that the velocity field of the narrow-line region
is dominated by the stellar gravitational potential (e.g., \citealt{bonningetal2005}).
However, the [OIII] line is known to often exhibit a blue wing that is associated with gas outflow and this can affect mass determinations \citep{marconietal2008}. 
Thus, \citet{komossaetal2008} fit the blue wing separately and use only the main [OIII] component to estimate $\sigma$.
This careful fitting can decrease the galaxy host mass measurement, placing the Narrow Line Seyfert 1 on the M$_{\rm BH}$-$\sigma$ relation \citep{komossaetal2008}.

Eight of the Peas are classified as NLS1s, which we define as 
galaxies with $500 \leq {\rm FWHM}_{\rm{H}\beta} \leq 2000~{\rm km}~{\rm s}^{-1}$.
We determined the black hole and galaxy masses for our sample of NLS1s.
Black hole masses are estimated using the relation given by  \citet{mclurejarvis2002}:
\begin{equation}
{\rm M}_{\rm BH}=10^{7.63} {\rm v}_{3000}^2 {\rm L}_{44}^{0.61}~M_\odot,
\end{equation}
 where v$_{3000}$ is the FWHM of the H$\beta$ line divided by 3000 km/s and $L_{44}$ is the luminosity at 5100\AA~divided by $10^{44} ~ \rm erg~s^{-1}$.
To determine the galaxy mass, we fit the [O III]  emission line with both a central narrow component and an additional blue wing.
We then take the central narrow component as a proxy for the stellar velocity dispersion $\sigma$.

\begin{figure}
\begin{center}
\includegraphics[angle=0, width=0.45\textwidth]{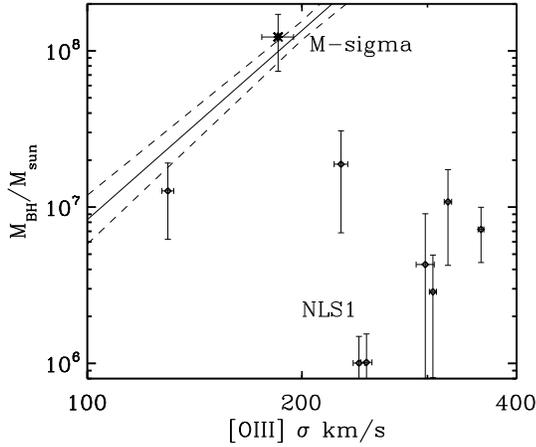}
 \caption{Using the [OIII] width as a measure of the host galaxy mass, we plot one broad-line AGN (star) and eight NLS1s (diamonds) on the ${\rm M_{BH}}$-$\sigma$ relation.  The NLS1s mostly lie far below the standard ${\rm M_{BH}}$-$\sigma$ relation, even though we determined galaxy mass from the narrow [OIII] component only.}
  \label{bhmass}
\end{center}
\end{figure}

Figure \ref{bhmass} shows the stellar velocity dispersion of the bulge, estimated from $\sigma_{\rm [O III]}$, vs black hole mass for the NLS1s, nearly all of which lie well below the classical M-$\sigma$ relation (solid line,  \citet{tremaineetal2002}; dashed lines are 1-$\sigma$ error contours)  even though we fit the [O III] line with an additional blue wing component.
This contrasts with a recent study that found this fitting method puts the NLS1s nearer to the M-$\sigma$ relation \citep{komossaetal2008}.
Instead, our results are consistent with studies measuring the galaxy mass using near-infrared bulge measurements \citep{ryanetal2007}.
We also note that single broad-line AGN found in our sample is consistent with the M-$\sigma$ relation.
Therefore the properties of our sample of NLS1s are consistent with some of those found in the literature, and can help in the study of the location of NLS1s on the M-$\sigma$ relation.

To better characterise the nuclear emission using high energy data, we searched the database at HEASARC \footnote{NASA's High Energy Astrophysics Science Archive Research Center; \texttt{http://heasarc.gsfc.nasa.gov/}}  for additional data on our sample.
Unfortunately, the Peas are well distributed throughout the 8,400 square degrees covered by the SDSS and not concentrated in any of the areas covered by deep multi-wavelength surveys.
None of the NLS1s are bright enough at soft X-ray wavelengths to be detected in the {\it ROSAT} All-Sky Survey \citep{vogesetal1999}.
We note that the broad-line AGN is detected with a luminosity of nearly $10^{44} \rm{ergs}$ $\rm{s^{-1}}$, as is one of the type 2 Seyferts with an X-ray luminosity of nearly $3\times10^{44} \rm{ergs}$ $\rm{s^{-1}}$ (0.2-2 keV). 
Given the redshift range of the Peas and the detection limits of the {\it Rosat} all sky survey, this limits the NLS1s' X-ray luminosity to below a few $\times 10^{44} \rm{ergs}$ $\rm{s^{-1}}$; however, this upper limit is still well within the typical range of Seyfert luminosities.

One of the HST images is of a NLS1 Pea (Figure \ref{hst}, bottom centre).  This galaxy looks distinctly different from the patchy HST images of the Peas powered by star formation.  It appears to be an edge on disk, with no sign of morphological disturbance.   This is consistent with what is seen in the morphologies of other samples of NLS1s  \citep{ryanetal2007}.

\begin{table*}
\caption{Narrow Line Seyfert 1s}
\label{tab:nls1}
\begin{tabular}{@{}lrrcrr}
\hline
SDSS ObjId & \multicolumn{1}{c}{RA [degs]} &  \multicolumn{1}{c}{Dec [degs]} &  z & \multicolumn{1}{c}{$\sigma_{\rm[OIII] }$ [${\rm km~s^{-1}}$]$^a$} & \multicolumn{1}{c}{M$_{\rm BH}$[$10^7M_\odot$]$^b$}  \\
\hline
\hline
       587726032799400204 &  211.917006 &    2.296711 & 0.30920 &  240.45   (4.47) &   0.100   (0.024) \\
       587733410983182549 &  214.828750 &   51.044473 & 0.32363 &  297.90   (8.69) &   0.429   (0.239) \\
       587733398646620415 &  245.539241 &   35.352080 & 0.26601 &  129.72   (2.47) &   1.270   (0.323) \\
       587731521734640128 &  117.387347 &   28.568545 & 0.33697 &  226.91   (4.96) &   1.883   (0.599) \\
       587731892187037787 &  172.279625 &   57.934812 & 0.31238 &  305.33   (3.50) &   0.287   (0.103) \\
       588017978351616137 &  171.563561 &   38.971510 & 0.33651 &  246.36   (4.38) &   0.102   (0.026) \\
       587739377230610665 &  124.500818 &   19.302802 & 0.32452 &  320.52   (3.83) &   1.081   (0.329) \\
       587739406266728813 &  239.238254 &   21.520959 & 0.23314 &  356.63   (3.31) &   0.720   (0.139) \\

\hline
\end{tabular}
\\$^a$  We measured the [OIII] line width fitting simultaneously for a blue wing and narrow Gaussian components.  The measurements reported here are for the narrow Gaussian component with 3 sigma measurement error in parentheses. 
\\$^b$  The black hole masses are determined as in Eqn 1, 1 sigma errors are in  parentheses. \\
\end{table*}

\section{The Properties of Star-Forming Peas}
\label{sec:sf}
From our spectral diagnostics we have 80 star-forming objects, with high signal-to-noise, which we look at in greater detail here.  They are listed in Table \ref{tab:sf}, where we include information from the SDSS DR7 archive in Columns 1-5: SDSS ObjId, RA, Dec, z and the equivalent width of [O III].  

\subsection{Star Formation Rates}
\label{sec:sfr}
\begin{figure}
\begin{center}
\includegraphics[angle=0, width=0.48\textwidth]{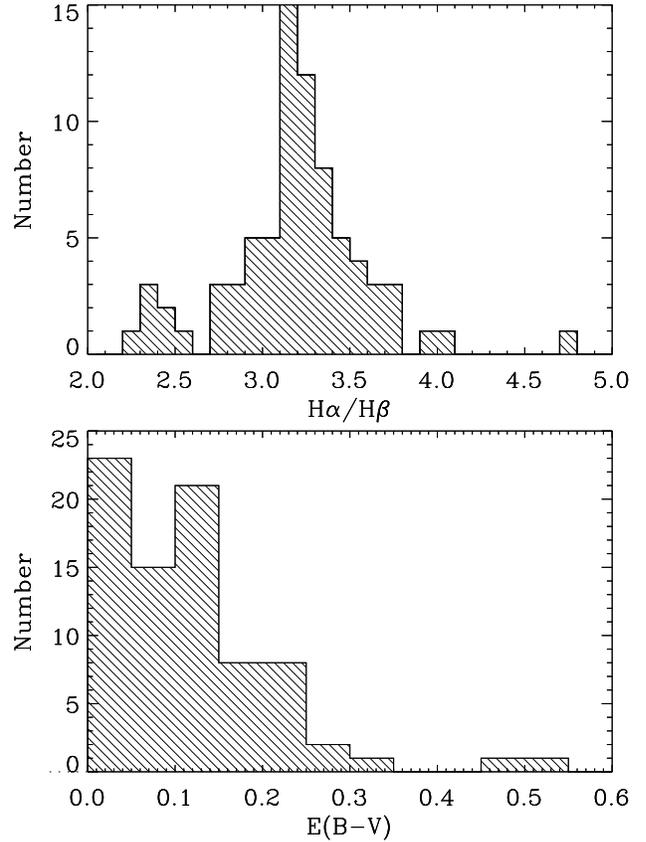}
 \caption{The histograms of H$\alpha$ / H$\beta$ (top) and the colour excess, $\rm E(B-V)$, (bottom) as
determined from the Balmer decrement. This distribution indicates that the
line-emitting regions of star-forming Peas are not highly reddened,
particularly compared to more typical star-forming or starburst galaxies.}
  \label{ebmv}
\end{center}
\end{figure}

To determine accurate star-formation rates using the H$\alpha$ fluxes, we corrected these recombination fluxes
 for both interestellar extinction and for the underlying stellar absorption lines in the stellar continuum \citep{kennicutt1998}. 
We measured the Balmer decrement, assuming an ${\rm R_V=A_V / E(B-V)}=3.1$, and using the \citet{cardelli1989} reddening curve and an intrinsic H$\alpha$ / H$\beta$ value of 2.85 (the Balmer decrement for case B recombination at T=$10^4$ K and $n_e=10^4$ cm$^{-3}$; \citealt{lequeux2005}).
There are a handful of star-forming Peas with H$\alpha$ / H$\beta$ less than 2.85, but these measurements are due to a combination of intrinsically low reddening and uncertainty in line flux determinations (Figure \ref{ebmv}, top panel); we set the extinction equal to zero in these cases.
Overall, we find that the reddening values for the Peas are low (Figure \ref{ebmv}, bottom panel), with nearly all Peas having ${\rm E(B-V) \leq 0.25}$.
Finally, using the corrected H$\alpha$ flux measurements, we measured star-formation rates \citep{kennicutt1998} up to $\sim30$ ${\rm M_{\odot}yr^{-1}}$.
The measured star formation rates are added to table \ref{tab:sf} in Column 7. 

\subsection{Stellar Mass}
\label{sec:mass}
The optical spectral energy distribution (SED) of the Peas is dominated by their
strong nebular emission lines. 
Thus we cannot apply standard SED fitting techniques directly to their photometric data. 
Additionally, virial masses are inaccessible  due to the low S/N of the spectral continuum and our inability to fit absorption lines.
We therefore turn to the SDSS spectra, where the emission lines can be subtracted or blocked out where necessary. 
The continuum of the resulting emission-line-free spectrum has very poor signal-to-noise, so
rather than fit this spectrum directly, we convolve it with a set of 19 medium-band
filters \citep{taniguchi2004} and treat the result as medium-band
photometric data. 
In other words, we construct an SED from the spectral continuum.
To this, we add the \textit{GALEX} near- and far-UV
photometric data points where available (\S~\ref{sec:luv}) and fit the SED with stellar population models.

\begin{figure*}
\begin{center}
\includegraphics[angle=90, width=0.98\textwidth]{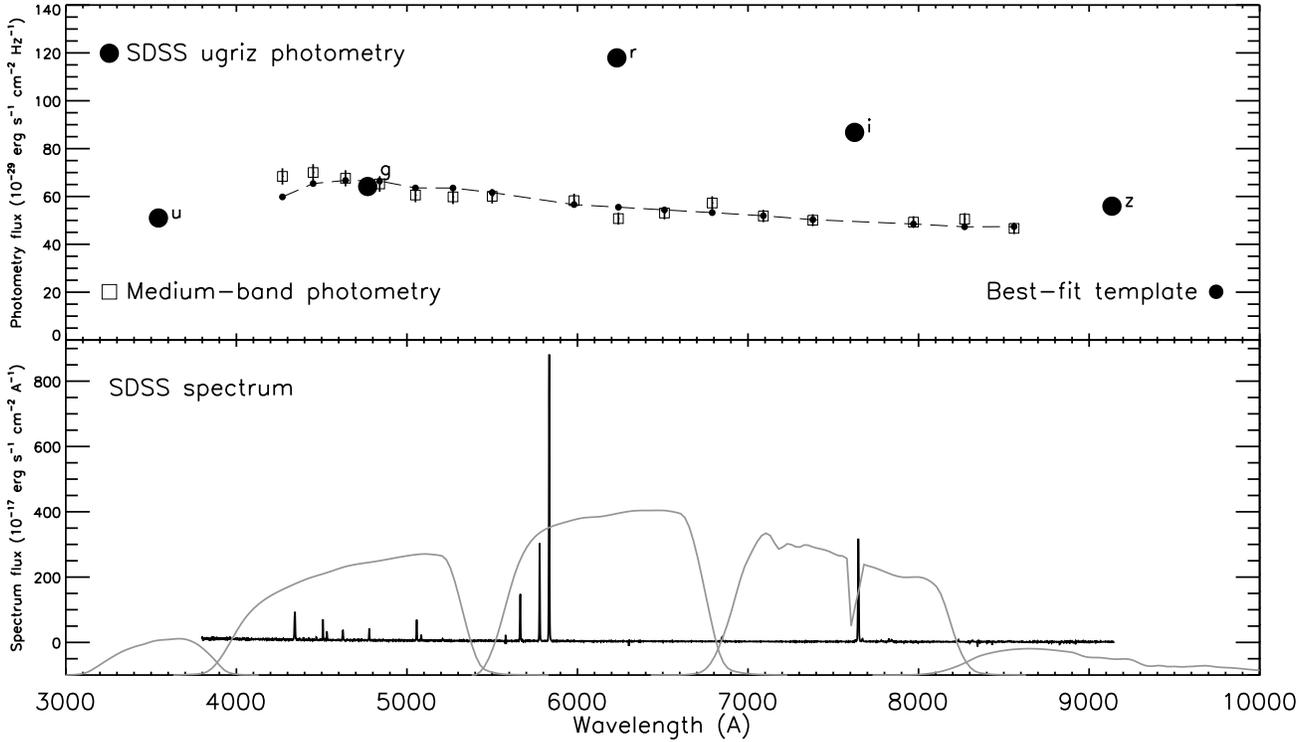}
 \caption{
 An example SED fit to a typical starforming Pea. The top panel shows both the
original emission-line dominated SDSS $ugriz$ photometry (large black points),
and the medium-band photometry derived from the emission-line subtracted
spectrum used for the SED fit (open boxes). The best-fit template (small black
points and dashed line) is plotted on top. In the bottom panel, we show the SDSS
spectrum of this Pea together with the $ugriz$ filter transmission curves. The
[OIII] line dominates the $r$-band flux and the $i$-band flux is significantly affected by H$\alpha$ emission. 
The SDSS broad-band photometry is
dominated by emission lines to such an extent that fitting stellar templates to
it cannot yield reliable results.
}
  \label{ksmass}
\end{center}
\end{figure*}

We employ a method similar to that of \citet{schawinskietal2007} and model the
star formation history with two bursts, to account for the possible presence of an
underlying old stellar population. 
We use the stellar models of \citet{marastonetal1998, maraston2005}
with the Salpeter IMF and a range of metallicities. 
We also account for dust
extinction following the \citet{calzettietal2000} law and fit the resulting model
photometry to the data and compute the $\chi^2$ statistic. 
After marginalising
over all parameters, we obtain a stellar mass and an estimate of its error.
Higher S/N spectra are required to constrain the mass ratio between with older and younger stellar components.
Therefore our systematic errors are much larger than the computed formal statistical errors.
To quantify this uncertainty, we compared our mass estimates using a second SED-fitting code \citep{krieketal2009}, again using a Salpeter IMF and \citet{marastonetal1998,maraston2005} models.
The second code fits a single model template and therefore measures only a young stellar population component, excluding the additional free parameter of a second older stellar population from \citet{schawinskietal2007}.
The single stellar population fits result in a total stellar mass 0.75 dex lower on average.
This result is not unexpected as a younger stellar population is more luminous and can account for the same amount of light with a smaller mass contribution.
Accounting for the average 0.75 offset between the two mass measurements, the residual dispersion was just under 0.3 dex.
Therefore each individual mass measurement is uncertain at this minimum level.
Because two stellar populations are more likely to reside in these galaxies (\S~\ref{sec:bcg}), we quote the masses from the two-burst model for the results in this paper acknowledging their uncertainty.  
Figure \ref{ksmass} shows an example stellar population fit.
The median stellar mass of a Pea in our sample $\sim10^{9.5} M_{\odot}$ and they range from $10^{8.5} M_{\odot}$  to nearly $10^{10.5} M_{\odot}$, indicating the Peas as a class are significantly less massive than an L$_{\star}$ galaxy.
Galaxy mass estimates are also included in Table \ref{tab:sf}, Column 9.

\subsection{Metallicity}
\label{sec:met}
\begin{figure}
\begin{center}
\includegraphics[angle=0, width=0.48\textwidth]{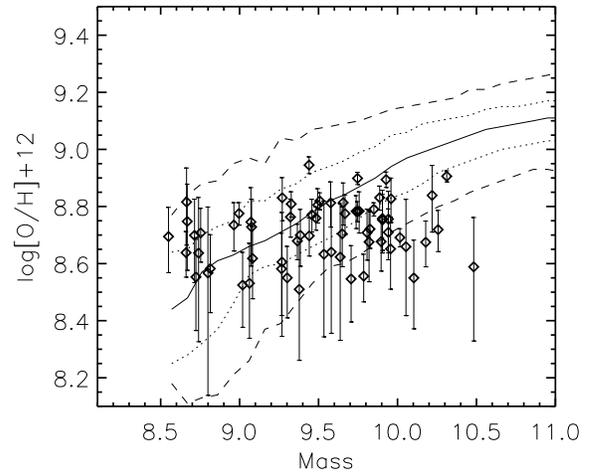}
 \caption{The mass metallicity relation after \citet{tremontietal2004}.  The Peas are sub-solar metallicity.  Although the Peas are in general consistent with the mass-metallicity relation, they depart from it at the highest mass end and thus do not follow the trend.  The Peas selection selects galaxies with a range of masses, but a more uniform metallicity.}
\label{met}
\end{center}
\end{figure}

We measure gas phase metallicity for the Peas using [N II] $\lambda$6584 \AA/ [O II] $\lambda\lambda$3726, 3729 \AA to estimate log[O/H] + 12 \citep{kewleydopita2002}.
The average Peas has a metallicity of  log[O/H] + 12 $\sim8.7$.
These metallicity measurements are broadly consistent with those determined from other line ratios for which we have lower (S/N) and fewer galaxies with measurements as well as with the location of the Peas on the BPT plot (Figure \ref{bpt}).
We plot the Peas on the mass metallicity relation from \citet{tremontietal2004} in Figure \ref{met}.
The lines are from \citet{tremontietal2004}, enclosing 68\% (dotted) and 95\% (dashed) of the star-forming galaxies in the  \citet{tremontietal2004} sample.
Although we do not see a trend in the Peas' metallicity with mass, they are roughly consistent with the mass-metallicity relation.
The exception to this agreement is for the Peas with the largest masses, which have the same low metallicity as their lower-mass counterparts and therefore lie below the mass-metallicity relation.
This  is likely due to the uncertainty in the mass determinations rather than the measurement of gas phase metallicity.
We include metallicity in Table \ref{tab:sf}, Column 8.

Overall, we find the Peas have  log[O/H] + 12 $\sim8.7$, sub-solar (Z$_\odot \sim$0.5) as per measurements of \citet{grevesseetal1998} and helioseismic measurements \citep{basuantia2008}, but near the solar abundances of \citet{asplundetal2005}.  These metallicities are common in low mass galaxies like the Peas.

\subsection{Specific Star Formation Rate}
\begin{figure}
\begin{center}
\includegraphics[angle=0, width=0.48\textwidth]{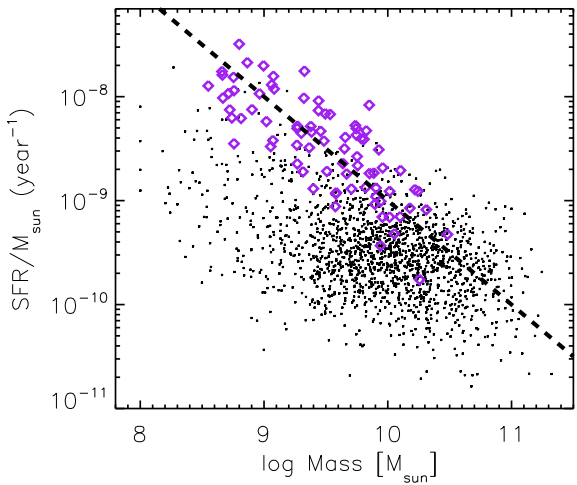}
 \caption{Specific star-formation rate vs. mass for the Peas (purple diamonds) and the Galaxy Zoo Mergers sample (black points).  The Peas have low masses typical of dwarf galaxies and much higher specific star formation rates compared to the merging galaxies.  The dashed line shows a constant star formation rate of 10 M$_{\odot}$yr$^{-1}$.  Most of the Peas had SFRs between 3 and 30 M$_{\odot}$yr$^{-1}$ and hence follow this line closely.}
  \label{ssfr}
\end{center}
\end{figure}

Combining the mass measurements with the star formation rates, we find extraordinarily high specific star formation rates for the Peas.
``Specific" star formation rate (SSFR) refers to the star formation rate per solar mass in units of year$^{-1}$ and thus can be directly related to the time taken to double the stellar mass of a system (1/SSFR).
In Figure \ref{ssfr} we compare the Peas to the Galaxy Zoo Merger Sample \citep{dargetal2009a}.  
Major mergers are frequently sites of active star formation, and yet the Peas are an order of magnitude higher in specific star formation rates than this comparison sample.
These specific star formation rates imply doubling times between 100 Myr and $\sim 1$ Gyr.
The uniformly high star forming rates of the Peas are not unexpected because their selection criteria targets strong emission lines.  
If the rates of star formation were lower, they would not be detected as ``green'' in the SDSS imaging.
However, the SSFR we measure are unusually high for galaxies at z$\sim$0.2, which typically reach SSFR$\sim10^{-9}$yr$^{-1}$ at most \citep{brinkmannetal2004,baueretal2005}.

\subsection{UV Luminosity}
\label{sec:luv}
The Peas are well spread throughout the 8,400 square degrees covered by the SDSS and not concentrated in any of the areas covered by deep multi-wavelength surveys.
However, {\it GALEX} GR4 is well matched to SDSS in depth and area and 139 of the 251 Peas are detected in {\it GALEX} GR4 data \citep{morrisseyetal2007}.
For the 57 of 80 star-forming Peas with Galex detections (S/N $\ge$ 3), the median luminosity is $3\times10^{10} L_{\odot}$.
We include L$_{\rm FUV}$ in Table \ref{tab:sf}, Column 6.
The very high UV luminosities combined with low reddening (\S \ref{sec:sfr}) are rare in local galaxies and are more typically found in galaxies at higher redshift.

\section{Comparison with other samples of galaxies}
\label{sec:comp}
\subsection{Blue Compact Dwarfs}
\label{sec:bcg}
Blue compact dwarfs are a subset of the local dwarf population, first identified by \citet{zwicky1965} as star-like field galaxies on Palomar Sky Survey Plates.
They are characterized by compact, gas-rich regions of high star formation rates  \citep{papaderosetal2002,vaduvescuetal2007}, often lying inside an older stellar population of larger spatial extent containing a significant mass contribution of the galaxy as a whole
\citep{looseetal1986,kunthetal1988,papaderosetal1996,aloisietal2007}.
They are generally low-metallicity (7.12 $\leq$ 12+log(O/H) $\leq$ 8.4), evolved, gas-rich dwarfs undergoing recurrent starburst activity \citep{looseetal1986, papaderosetal1996, gildepaz2005, papaderosetal2008}.
As a class Blue Compact Dwarfs cover a wide range in absolute magnitude and are often sub-classified by size (Ultra Compact Blue Dwarfs) or luminosity (Luminous Blue Compact Galaxies).

Ultra Compact Blue Dwarf galaxies are the smallest of the Blue Compact Dwarfs with physical diameters less than 1 kpc and low masses ($\sim$ $10^7-10^9$M$_{\odot}$; \citealt{corbinetal2006}).
Typically an order of magnitude less luminous than other BCGs, their optical spectra can be dominated by very strong emission lines  \citep{gusevaetal2004,corbinetal2006} and they can contain substantial amounts of internal dust (eg. $E(B-V)\sim0.28$ ;   \citealt{corbinvacca2002}).
Like other low mass galaxies, Ultra Blue Compact Dwarfs have very low metallicities, yet they still lie on an extrapolation of the \citet{tremontietal2004} mass-metallicity relation \citep{corbinetal2006}.
These galaxies can show asymmetric morphologies with multiple sub-clumps of active star-forming regions \citep{corbinvacca2002, corbinetal2006}.
While their light is dominated  by the compact, young ($\sim$1-10 Myr) stellar population, their stellar mass is dominated by an older, evolved ($\sim$10 Gyr) population.
As a population, these galaxies tend to reside within voids \citep{corbinetal2006}.

At the other end of the Blue Compact Dwarf category, Luminous Blue Compact Galaxies are the most luminous members with $M_B \leq -17.5$.
They are more massive than their Ultra Compact counterparts, with average masses near 5$\times$10$^9$ M$_{\odot}$ \citep{guzmanetal2003}.
As one might expect from a class of more massive galaxies, their metallicities are on average higher than the Ultra Blue Compact Dwarfs  (7.7 $\leq$ 12+log(O/H) $\leq$ 8.4; \citealt{gusevaetal2004,hoyosetal2007}).
However, like their Ultra Compact Blue Dwarf counterparts, Luminous Blue Compact Galaxies show compact prominent regions of star formation, often with disturbed morphologies \citep{bergvalljohansson1985} and star formation rates  typically ranging from 1-5 M${\rm _{\odot} yr^{-1}}$.
Underlying these young, star-forming regions, older stellar populations with the colours and stellar profiles of older, massive ellipticals are detected in near-IR imaging  \citep{bergvallostlin2002}.
Examples of Luminous Blue Compact Galaxies are found at intermediate redshifts  \citep{guzmanetal1997,phillipsetal1997,hoyosetal2004} and may contribute up to 50\% of the star formation rate density in the Universe at z=1 \citep{guzmanetal1997}.

Because both the Peas and Blue Compact Dwarf galaxies contain strong emission lines originating in compact regions of star formation, we investigated their potential overlap as a class.
Although the Peas are similar in morphology, environment and physical size to the z=0 Ultra Blue Compact Dwarfs, they appear to be a different class of galaxies.
The Peas have significantly higher metallicities, typically more than 0.5 Z$_\odot$ compared to 0.02 Z$_\odot$ for the Ultra Compact Blue Dwarfs.
Additionally, the stellar masses of the Peas are on average $\sim$10$^{9.5}$M$_\odot$, roughly an order of magnitude larger than similar measurements of Ultra Compact Blue Dwarfs \citep{corbinetal2006}.
While Ultra Compact Blue Dwarfs are found near $z\sim0$, the Peas are detected at 0.113$\leq$z$\leq$0.36, meaning they have several Gyrs to grow and further increase their metallicity before reaching z=0, increasing the disparity in the measured masses and metallicities.
In contrast, the largest of the Blue Compact Dwarfs, Luminous Blue Compact Galaxies have stellar masses and metallicities that match the measured values for the Peas.
The luminosities of the Peas are easily as bright as this class of galaxies, with median $M_B\sim-20$.
Additionally, studies of low-metallicity SDSS objects like Blue Compact Dwarfs find that they lie in the same location on the BPT plot  shown in Figure \ref{bpt} \citep{izotovetal2006} and [O III] 5007 / H$\beta$ line ratios for a sample of four Luminous Compact Blue Galaxies measured with STIS ($\log([O III] \lambda 5007 \AA~H\beta) \sim 0.5$; \citealt{hoyosetal2004}) are similar to those we find here (Figure \ref{bpt}).
Studies of the Peas at near-infrared wavelengths could potentially reveal older stellar populations, like those found in Luminous Blue Compact Galaxies.  
These older stellar populations could reveal additional stellar mass and potentially larger radii.
Therefore, the Peas could potentially be classified as part of the heterogeneous Luminous Blue Compact Galaxies category.

We tentatively conclude that the Peas form a different class of galaxies than Ultra Blue Compact Dwarfs, but they are similar to the most luminous members of the Blue Compact Dwarfs category, Luminous Blue Compact Galaxies.

\subsection{Local UV-Luminous Galaxies}
\label{sec:uvlg}
\citet{heckmanetal2005} defined a sample of 215 ultraviolet-luminous galaxies (UVLGs) from the SDSS spectroscopic galaxy sample (DR3) and {\it GALEX} (GR1) with luminosities $L_{\rm FUV} \ge 2\times10^{10} L_{\odot}$.
We have 139 Peas with Galex detections (S/N $\ge$ 3), 44 of which meet \citet{heckmanetal2005}'s criteria ($L_{FUV} \ge 2\times10^{10} L_{\odot}$, z$\leq$0.3 and SDSS pipeline spectral type identified as \texttt{galaxy}.
Since GR4+DR7 cover a much larger area than GR1+DR3, we find only four overlapping sources between the Peas and the UVLG sample \citep{hoopesetal2007}.
However, we note that none of the other 211 UVLGs fall into the SDSS colour-selection wedge of the Peas.

\begin{figure}
\begin{center}
\includegraphics[angle=0, width=0.48\textwidth]{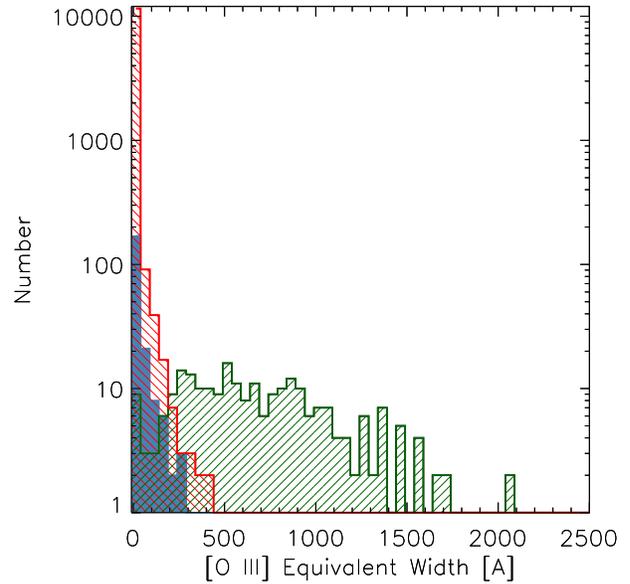}
 \caption{Neither the distribution of [O III] equivalent width for $\sim10,0000$ comparison galaxies (\S \ref{sec:sample}; red lined histogram) nor the UVLG sample (blue solid histogram) have equivalent widths comparable to the Peas (green lined histogram).  As a class the Peas have much larger [O III] equivalent widths than normal or UV-luminous galaxies.  All equivalent widths displayed here are from the SDSS pipeline measurements.}
  \label{ew}
\end{center}
\end{figure}

The Peas selection includes many UV-luminous galaxies, but does not uncover the same population of galaxies as a selection based on UV luminosity.
To illustrate this in Figure \ref{ew}, we compare the [O III] equivalent width, as measured by the SDSS pipeline, for the galaxy sample described in Section \ref{sec:sample} and the \citet{heckmanetal2005} UVLG sample to that of the Peas.
Unsurprisingly, due to their selection, the Peas have much larger equivalent width measurements than either the UVLG sample or the general spectroscopic galaxy sample (\S~\ref{ew}).
The Pea [O III] equivalent widths are due to the  combination of both very high star formation rates and the faint continuum.
The Peas form a sample of objects selected to have large [O III] equivalent widths, many of which also have large UV luminosities.
As the UVLGs are currently one of the best local samples of galaxies analogous to high redshift Lyman-break galaxies \citep{hoopesetal2007}, we now look into the comparison between the Peas and high redshift galaxy samples.

\subsection{UV-luminous High Redshift Galaxies}
\label{sec:lbgs}
Peas are sites of extreme star-formation in the local Universe.
They are both very compact and have low stellar masses, yet they have enormous star formation rates, as shown by their emission lines and their UV luminosity.
This sample of galaxies is quite distinct from typical $z\sim0$ star-forming galaxies, i.e., blue spiral galaxies or dusty irregulars \citep{kennicutt1998}.  
Therefore, we look at the properties of higher redshift galaxies where the bulk of the star formation in the Universe is occurring \citep{madauetal1996,lillyetal1996,reddyetal2009}.

Various colour-selection techniques have been developed for selecting samples of high redshift galaxies, including the Lyman-break dropout technique, which selects $z\sim3$ galaxies by identifying $U$-band dropouts in deep imaging (e.g., \citealt{steideletal2003}), and Ly$\alpha$ emission selection, which uses narrow-band imaging to select galaxies at various redshifts due to the excess emission of Ly$\alpha$ compared to the neighbouring continuum \citep{cowieandhu1998, thommesetal1998, kudritzkietal2000,steideletal2000}.
Both techniques return a similar population of galaxies, but Ly$\alpha$ emitters are characterised by a fainter continuum and stronger equivalent widths in their emission lines \citep{giavaliscoetal2002}.
Although both selection techniques have different biases, the galaxies showing Ly$\alpha$ in emission tend to be smaller and younger on average \citep{gawiseretal2007, finkelsteinetal2007}.
UV-luminous high redshift galaxies, like Ly-$\alpha$ emitters and Lyman-break galaxies, are characterized by a high UV luminosity and a relatively low obscuration by dust \citep{gawiseretal2006lya,venemansetal2005}.
They are starburst galaxies with strong blue continua dominated by young, massive stars \citep{giavaliscoetal2002}.
These techniques have been extended to other redshifts out to $z\sim6$, using different photometric bands to select drop-out galaxies (e.g., \citealt{ouchietal2001,stanwayetal2003,giavaliscoetal2004,kakazuetal2007}).
In a broad sense, the Peas are an application of a Ly$\alpha$-like selection technique at much lower redshift and targeting the [O III] emission line.

UV-luminous, high-redshift galaxies are similar in size to the Peas, with half light radii $\lesssim$ 2 kpc \citep{giavaliscoetal1996,bremeretal2004,pascarelleetal1998}.
Many of these high redshift galaxies show disturbed morphologies  \citep{ravindranathetal2006,lotzetal2006}, like those seen in the HST images of the star-forming Peas in Figure \ref{hst}.
In Lyman-break galaxies, rest-frame equivalent widths of [O III] can reach up to hundreds of \AA~ \citep{pettinietal2001}, although most are much lower.
For Ly$\alpha$ emitters, most are selected to have EW of Ly$\alpha$ $\ge$ 20 \AA~and can range up to $\sim$240 \AA~ or more \citep{gawiseretal2006lya, gronwalletal2007}.
In this way, the Peas are more like Ly$\alpha$ emitters, selected for their strong emission lines, and the subset of Lyman-break galaxies with large emission lines.
Reddening measures of Lyman-break galaxies are low, similar to those we measure for the Peas, ${\rm E(B-V)} \lesssim 0.2$ \citep{shapleyetal2003,giavaliscoetal2002}.  
The Ly$\alpha$ line is also easily suppressed by dust, making Ly$\alpha$ emitters similarly low in dust \citep{gawiseretal2006lya}.
Additionally they have high star formation rates of a few to tens of solar masses per year \citep{barmbyetal2004,lehmeretal2005lbg,coppinetal2007, carillietal2008,mannuccietal2009, pentericcietal2009}, again similar to what we measure for the Peas.
Peas are similar to UV-luminous high redshift galaxies in size, morphology, large emission lines, reddening and star formation rate.

Masses measured for high redshift UV-luminous galaxies are slightly larger than those for the average Pea, $\sim10^9 M_\odot - 10^{11} M_\odot$ \citep{barmbyetal2004, coppinetal2007, yabeetal2008, mannuccietal2009,pentericcietal2009}.
In terms of SSFR vs mass, Lyman Alpha Emitters  lie just below the 10 M$_\odot$yr$^{-1}$ line (Figure 12, dashed line) \citep{gawiseretal2006lya, castroceronetal2006}.
While LBGs range from 10 M$_\odot$yr$^{-1}$ to 100 M$_\odot$yr$^{-1}$, at masses below $10^9.5$ they are preferentially near 100M$_\odot$yr$^{-1}$ \citep{castroceronetal2006,shapleyetal2001,barmbyetal2004}.
Metallicities of Lyman-break galaxies are typically measured near 10-50\% solar \citep{pettinietal2001,mannuccietal2009}, significantly lower than that of the lower redshift Peas.
This is not surprising since the Peas have had significantly longer to enrich their gas.
Finally, Lyman-break galaxies are known to be strongly clustered and found in the densest regions \citep{giavaliscoetal2002} in contrast to the Peas.
The peas are found in lower density regions, have lower masses and smaller specific star formation rates than the UV-luminous galaxies found at high redshift.
The smaller mass and lower density environment of the Peas is consistent with a picture of downsizing \citep{cowieetal1996,thomasetal2005} where smaller present day galaxies form their stars at later times in lower density environments.

Understanding the evolution of starburst galaxies over cosmic time is central to understanding the build up of stars in galaxies.
At high redshift ($1.9\leq z\leq3.4$), UV-luminous galaxies are responsible for the formation of a large fraction ($\sim 40\%$) of the present day stellar mass \citep{reddyetal2009}.
Therefore, the Peas are potentially the remnants of a mode of star formation that was common in the early Universe. 
If that is the case, then the Peas are an ideal laboratory for understanding this mode of star formation, as
their continuum properties are easily accessible with large ground-based telescopes.
Additionally, their low redshifts allow optical and near-infrared telescopes to investigate potential underlying older stellar mass components.
High redshift galaxies are both small and faint, even at HST resolution, but the galaxies at $z\sim 0.1-0.3$ can be imaged at higher physical resolution \citep{overzieretal2008}.
Studies of the morphologies of low dust, high star-forming galaxies in the local Universe can lend insight to the processes occurring at higher redshift where the morphologies cannot be as finely resolved.
The X-ray luminosites of known Lyman-break galaxies are $\sim10^{41}$ ergs/s, accessible at z$\sim$0.2 with snapshots from {\it Chandra} \citep{Hornschemeieretal2008}, allowing for studies of their star formation rates at X-ray wavelengths.
The Peas may be the last remnants of a mode of star formation common in the early Universe, and therefore an excellent laboratory for understanding that mode.

\section{Summary}
We investigated a class of galaxies, known as Peas, discovered by the Galaxy Zoo project.  
These galaxies are characterised by a distinctly green colour in $gri$ imaging arising from a very large [O III] equivalent width.  
\begin{itemize}
\item{251 Peas were collected from the SDSS spectroscopic database based on a colour selection in the redshift range $0.112 \leq z \leq 0.360$.}
\item{The Peas are unresolved in SDSS imaging, placing an upper limit on their physical radius of approximately 5 kpc.}
\item{The median environmental density around the Peas is less than two-thirds of that around normal galaxies.}
\item{The BPT spectral line diagnostic reveals that the majority of the Peas are star-forming galaxies, some of which show patchy morphology in HST imaging.}
\item{We uncover 8 new Narrow Line Seyfert 1s from the SDSS archive.  They lie below the M-$\sigma$ relation, similar to other samples of NLS1s. }
\item{From a sample of 80 star-forming galaxies with high S/N spectral measurements, we find that the Peas have very large star formation rates (up to 30 M$_\odot$yr$^{-1}$), low stellar mass ($\sim10^{9.5}$ M$_\odot$), low metallicity (log[O/H]+12$\sim$8.7) and large UV luminosities ($\sim 3\times10^{10}$L$_\odot$).}
\item{The Peas form a different class of galaxies than Ultra Blue Compact Dwarfs, but may be similar to the most luminous members of the Blue Compact Dwarfs category.  Luminous Blue Compact Galaxies are similar to the Peas in their masses, morphologies, metallicities, luminosities and redshifts.  It would be interesting to study the Peas at NIR wavelengths to see if they have older underlying stellar populations like those found in Luminous Blue Compact Galaxies.}
\item{The Peas share properties similar to local UV-selected samples in Sloan, but uncover a different population with more extreme equivalent widths of [O III] emission line.}
\item{The Peas are similar to UV-luminous high redshift galaxies such as Lyman-break Galaxies and Ly$\alpha$ emitters.  However, these high redshift galaxies are higher in mass, lower in metallicity and found in the densest regions.  The smaller mass and lower density environment of the Peas is consistent with a picture of downsizing, where smaller present day galaxies form their stars at later times in lower density environments.  If the underlying processes occurring in the Peas is similar to that found in the UV-luminous high redshift galaxies, the Peas may be the last remnants of a mode of star formation common in the early  Universe.}
\end{itemize}

\section{Acknowledgements}
We wish to thank the ``Peascorps'' for all their hard work: including, Elisabeth Baeten, Gemma Coughlin, Dan Goldstein, Brian Legg, Mark McCallum, Christian Manteuffel, Richard Nowell, Richard Proctor, Alice Sheppard, Hanny van Arkel and Alice Sheppard for her help gathering information about the discovery of the Peas in Galaxy Zoo.  
We also thank Bethany Cobb for help with the editing of this paper.
We would also like to thank Eric Gawiser, Pieter van Dokkum, Erin Bonning, Soo Lee, Sarbani Basu and Sugata Kaviraj for helpful comments and our anonymous referee for constructive and knowledgeable comments which have contributed to the discussion.

Support from NSF grant \#AST0407295 and Yale University is gratefully acknowledged.
CJL acknowledges support from the STFC
Science in Society Programme. 

Funding for the SDSS and SDSS-II has been provided by the Alfred
P. Sloan Foundation, the Participating Institutions, the National
Science Foundation, the U.S. Department of Energy, the National
Aeronautics and Space Administration, the Japanese Monbukagakusho, the
Max Planck Society, and the Higher Education Funding Council for
England. The SDSS Web Site is http://www.sdss.org/.

The SDSS is managed by the Astrophysical Research Consortium for the
Participating Institutions. The Participating Institutions are the
American Museum of Natural History, Astrophysical Institute Potsdam,
University of Basel, University of Cambridge, Case Western Reserve
University, University of Chicago, Drexel University, Fermilab, the
Institute for Advanced Study, the Japan Participation Group, Johns
Hopkins University, the Joint Institute for Nuclear Astrophysics, the
Kavli Institute for Particle Astrophysics and Cosmology, the Korean
Scientist Group, the Chinese Academy of Sciences (LAMOST), Los Alamos
National Laboratory, the Max-Planck-Institute for Astronomy (MPIA),
the Max-Planck-Institute for Astrophysics (MPA), New Mexico State
University, Ohio State University, University of Pittsburgh,
University of Portsmouth, Princeton University, the United States
Naval Observatory, and the University of Washington.

\clearpage
\begin{deluxetable}{lrrlrccccccc}
\tabletypesize{\scriptsize}
\tablecolumns{11}
\tablewidth{0pc}
\tablecaption{Properties of Pea Star-Forming Galaxies}
%\begin{table*}
%\caption{Star Forming Galaxies}
%\begin{tabular}{@{}lrrlccccccc}
\tablehead{
\colhead{SDSSID$^{\rm a}$} & \colhead{RA$^{\rm a}$} & \colhead{Dec$^{\rm a}$} &  \colhead{z$^{\rm a}$} &  \colhead{[O III] EW$^{\rm a}$} & \multicolumn{2}{c}{$L_{\rm FUV}$$^{\rm b}$} &   \multicolumn{2}{c}{SFR $^{\rm c}$} & \multicolumn{2}{c}{log(O/H)+12$^{\rm d}$} & \colhead{Stellar Mass$^{\rm e}$} \\
\colhead{} & \colhead{(J2000)} & \colhead{(J2000)} & \colhead{} & \colhead{\AA} & \multicolumn{2}{c}{[$10^{44}$ ergs ${\rm s^{-1}}$]}  &   \multicolumn{2}{c}{[M$_\odot {\rm yr^{-1}}$]} &  \multicolumn{2}{c}{} & \colhead{[M$_\odot$]} 
}
\startdata
              587725073921409255 &  146.242618 &   -0.762639 &  0.3002 &   312.8 &    1.04&( 0.21) &   12.45&(0.52) &    8.78& $^{8.85 }_{8.71}$ &    9.76  \\
       588848899919446344 &  195.546460 &   -0.087897 &  0.2255 &   834.4 &    0.19&( 0.04) &    4.47&(0.22) &    8.75& $^{8.87 }_{8.59}$ &    9.07  \\
       587725576962244831 &  261.776373 &   59.817273 &  0.3472 &   651.5 &  \nodata&(\nodata) &   24.98&(1.48) &    8.71& $^{8.79 }_{8.61}$ &    9.81  \\
       587731187273892048 &  351.413453 &    0.752012 &  0.2770 &   319.6 &    0.99&( 0.06) &   11.11&(0.85) &    8.70& $^{8.79 }_{8.59}$ &    9.38  \\
       587731513693503653 &   50.687082 &    0.745111 &  0.3043 &   282.1 &    1.09&( 0.27) &   14.22&(0.80) &    8.83& $^{8.87 }_{8.79}$ &    9.89  \\
       587724233716596882 &   22.292299 &   14.992956 &  0.2800 &   354.0 &    1.53&( 0.21) &   13.37&(0.72) &    8.77& $^{8.83 }_{8.71}$ &    9.46  \\
       587727179006148758 &   45.839226 &   -7.989791 &  0.1650 &   886.0 &  \nodata&(\nodata) &    8.72&(0.64) &    8.71& $^{8.79 }_{8.61}$ &    8.75  \\
       587724241767825591 &   51.556792 &   -6.586816 &  0.1621 &   822.9 &  \nodata&(\nodata) &   11.44&(0.41) &    8.76& $^{8.80 }_{8.72}$ &    9.48  \\
       587724240158589061 &   54.949128 &   -7.428132 &  0.2608 &   397.6 &    2.00&( 0.14) &   28.96&(1.68) &    8.78& $^{8.84 }_{8.72}$ &    9.74  \\
       587726032778559604 &  164.319700 &    2.535293 &  0.3028 &   348.5 &    1.35&( 0.14) &    8.75&(0.57) &    8.76& $^{8.86 }_{8.64}$ &    9.90  \\
       587726032253419628 &  191.097382 &    2.261231 &  0.2395 &  1151.4 &  \nodata&(\nodata) &   25.24&(1.30) &    8.70& $^{8.76 }_{8.63}$ &    9.44  \\
       588010360138367359 &  130.570630 &    3.635203 &  0.2194 &   567.5 &  \nodata&(\nodata) &    6.60&(0.21) &    8.55& $^{8.66 }_{8.40}$ &    9.71  \\
       587726102030451047 &  236.787938 &    3.603914 &  0.2314 &   891.8 &  \nodata&(\nodata) &    9.83&(0.61) &    8.74& $^{8.81 }_{8.64}$ &    8.96  \\
       587729155743875234 &  173.265848 &   65.228162 &  0.2414 &   475.2 &    1.61&( 0.18) &    5.43&(0.24) &    8.66& $^{8.83 }_{8.42}$ &   10.05  \\
       587728919520608387 &  212.938906 &   62.653138 &  0.2301 &   529.0 &    1.45&( 0.16) &   12.80&(0.47) &    8.67& $^{8.75 }_{8.59}$ &   10.18  \\
       587729229297090692 &  234.405309 &   58.794575 &  0.2143 &   851.6 &    0.44&( 0.06) &    6.32&(0.20) &    8.58& $^{8.71 }_{8.42}$ &    9.27  \\
       587725818034913419 &  235.209139 &   57.411652 &  0.2944 &   173.3 &    3.50&( 0.24) &   16.81&(1.18) &    8.91& $^{8.92 }_{8.89}$ &   10.31  \\
       587730774416883967 &  339.396081 &   13.613062 &  0.2936 &   438.3 &  \nodata&(\nodata) &   24.78&(0.86) &    8.55& $^{8.68 }_{8.37}$ &   10.10  \\
       587730774965354630 &    6.716985 &   15.460460 &  0.2136 &   754.2 &  \nodata&(\nodata) &    3.32&(0.21) &    8.81& $^{8.89 }_{8.73}$ &    9.58  \\
       587728906099687546 &  117.403215 &   33.621219 &  0.2733 &   339.2 &    2.04&( 0.26) &   58.83&(2.61) &    8.79& $^{8.81 }_{8.77}$ &    9.85  \\
       587725550133444775 &  156.563375 &   63.552363 &  0.3338 &   933.3 &    1.27&( 0.28) &    3.76&(0.28) &  \nodata& $^{... }_{...}$ &    9.05  \\
       588009371762098262 &  170.582224 &   61.912629 &  0.2045 &   960.7 &    0.40&( 0.07) &    5.59&(0.23) &    8.70& $^{8.83 }_{8.53}$ &    8.71  \\
       588011122502336742 &  181.772142 &   61.586621 &  0.2620 &   760.2 &    0.68&( 0.13) &   13.02&(0.62) &  \nodata& $^{... }_{...}$ &    9.85  \\
       588011103712706632 &  226.616617 &   56.450741 &  0.2786 &   341.9 &    1.08&( 0.17) &   23.88&(0.93) &    8.90& $^{8.92 }_{8.88}$ &    9.75  \\
       588013384341913605 &  141.501678 &   44.460044 &  0.1807 &   651.5 &    1.55&( 0.11) &   14.35&(0.89) &    8.62& $^{8.73 }_{8.48}$ &    9.08  \\
       587732134315425958 &  195.368010 &   51.080893 &  0.3479 &   554.4 &    3.20&( 0.26) &   23.17&(1.74) &    8.63& $^{8.82 }_{8.34}$ &    9.53  \\
       587729777439801619 &  204.299529 &   -2.434842 &  0.2737 &   399.2 &  \nodata&(\nodata) &    8.63&(0.35) &    8.76& $^{8.85 }_{8.64}$ &    9.94  \\
       587729777446945029 &  220.630713 &   -2.164466 &  0.2938 &  1456.6 &    1.61&( 0.19) &   20.22&(0.80) &    8.57& $^{8.80 }_{8.14}$ &    8.80  \\
       587732152555864324 &  116.991682 &   23.609113 &  0.1552 &   874.9 &  \nodata&(\nodata) &    3.26&(0.07) &  \nodata& $^{... }_{...}$ &    9.40  \\
       587732578845786234 &  157.912214 &    7.265701 &  0.2525 &   763.9 &    0.57&( 0.18) &    6.04&(0.20) &    8.52& $^{8.64 }_{8.38}$ &    9.02  \\
       587733080270569500 &  163.378431 &   52.631353 &  0.2526 &   418.2 &    3.61&( 0.26) &   27.59&(1.09) &    8.78& $^{8.82 }_{8.75}$ &    9.75  \\
       588297864714387604 &  131.975356 &   33.615227 &  0.3063 &   323.7 &    1.89&( 0.20) &   21.46&(0.71) &    8.81& $^{8.86 }_{8.74}$ &    9.50  \\
       587735695911747673 &  204.919632 &   55.461137 &  0.2291 &    42.1 &    1.41&( 0.16) &    3.22&(0.13) &    8.71& $^{8.79 }_{8.61}$ &    9.94  \\
       587735696987717870 &  213.630037 &   54.515587 &  0.2270 &   773.7 &  \nodata&(\nodata) &    4.04&(0.10) &    8.58& $^{8.70 }_{8.43}$ &    8.81  \\
       587733441055359356 &  251.527242 &   31.514859 &  0.2907 &   868.4 &    1.20&( 0.14) &    6.56&(0.36) &  \nodata& $^{... }_{...}$ &    8.76  \\
       588017605211390138 &  154.513517 &   41.105860 &  0.2371 &  1191.4 &    0.88&( 0.14) &    8.83&(0.25) &    8.68& $^{8.79 }_{8.54}$ &    9.82  \\
       588017114517536797 &  216.023868 &   42.279524 &  0.1848 &  1348.8 &    1.49&( 0.12) &   19.66&(1.09) &    8.78& $^{8.81 }_{8.73}$ &    9.00  \\
       588017116132540589 &  228.535985 &   38.868716 &  0.3324 &   632.5 &    1.77&( 0.22) &    6.02&(0.36) &  \nodata& $^{... }_{...}$ &    8.90  \\
       588018090541842668 &  235.755108 &   34.767079 &  0.1875 &   673.9 &  \nodata&(\nodata) &    2.02&(0.05) &  \nodata& $^{... }_{...}$ &    8.76  \\
       588018090013098618 &  251.898063 &   22.783002 &  0.3138 &   578.8 &  \nodata&(\nodata) &    4.20&(0.49) &    8.61& $^{8.78 }_{8.35}$ &    9.27  \\
       588016878295515268 &  137.879799 &   31.457439 &  0.2718 &   426.3 &    1.05&( 0.12) &    9.18&(0.37) &  \nodata& $^{... }_{...}$ &    9.27  \\
       587735661007863875 &  139.260529 &   31.872384 &  0.3002 &   219.7 &    2.08&( 0.18) &   20.29&(1.51) &    8.95& $^{8.97 }_{8.92}$ &    9.44  \\
       588016892783820948 &  148.712329 &   37.365500 &  0.2834 &   279.9 &    0.90&( 0.16) &    8.42&(0.50) &    8.78& $^{8.84 }_{8.70}$ &    9.67  \\
       587735663159738526 &  149.415718 &   37.702114 &  0.2867 &   235.3 &    1.96&( 0.23) &   15.02&(0.63) &  \nodata& $^{... }_{...}$ &    9.75  \\
       588018055114784812 &  220.041419 &   46.326930 &  0.3008 &   304.4 &    3.17&( 0.26) &   31.50&(2.70) &    8.72& $^{8.78 }_{8.65}$ &    9.83  \\
       588018055652769997 &  223.648271 &   45.482288 &  0.2687 &   463.0 &    1.23&( 0.13) &   21.17&(2.67) &    8.84& $^{8.94 }_{8.71}$ &   10.22  \\
       588017570848768137 &  192.144310 &   12.567480 &  0.2634 &  1055.8 &    1.30&( 0.04) &   15.01&(0.75) &    8.53& $^{8.67 }_{8.34}$ &    9.06  \\
       587736915687964980 &  241.152768 &    8.333082 &  0.3123 &  1388.3 &    0.55&( 0.13) &   10.64&(0.58) &    8.75& $^{8.84 }_{8.65}$ &    9.90  \\
       587736915687375248 &  239.858241 &    8.688655 &  0.2970 &   904.8 &    0.66&( 0.13) &    3.44&(0.38) &    8.64& $^{8.83 }_{8.33}$ &    8.74  \\
       587738410863493299 &  152.987850 &   13.139471 &  0.1439 &  2388.3 &  \nodata&(\nodata) &    7.99&(0.22) &    8.64& $^{8.71 }_{8.55}$ &    8.66  \\
       587735349111947338 &  184.766599 &   15.435698 &  0.1957 &  1487.9 &  \nodata&(\nodata) &    7.47&(4.83) &    8.82& $^{8.93 }_{8.67}$ &    8.66  \\
       587738570859413642 &  204.867933 &   15.278369 &  0.1921 &  1289.0 &    0.64&( 0.06) &   18.81&(1.57) &    8.83& $^{8.90 }_{8.74}$ &    9.96  \\
       587736940372361382 &  217.614622 &   34.154720 &  0.1911 &   586.8 &    0.31&( 0.06) &    3.92&(0.15) &  \nodata& $^{... }_{...}$ &    9.31  \\
       587739153352229578 &  117.990764 &   16.637010 &  0.2647 &   440.1 &    0.66&( 0.13) &    6.29&(0.21) &    8.65& $^{8.76 }_{8.51}$ &    9.96  \\
       587738947196944678 &  123.966679 &   21.939902 &  0.1410 &  1582.0 &    0.27&( 0.03) &    3.96&(0.05) &    8.55& $^{8.69 }_{8.37}$ &    8.72  \\
       587738371672178952 &  125.698590 &   22.695578 &  0.2163 &  1040.6 &    0.90&( 0.13) &   37.41&(4.15) &    8.81& $^{8.85 }_{8.76}$ &    9.33  \\
       588017978880950451 &  150.556494 &   34.704908 &  0.3210 &   222.1 &  \nodata&(\nodata) &    4.43&(0.26) &    8.64& $^{8.82 }_{8.36}$ &    9.58  \\
       587739408388980778 &  174.342249 &   35.407413 &  0.1945 &   647.2 &    1.88&( 0.08) &   20.38&(0.91) &    8.76& $^{8.83 }_{8.69}$ &    9.32  \\
       588017977277874181 &  171.657352 &   38.050810 &  0.2469 &   646.9 &    1.56&( 0.11) &   24.18&(0.64) &    8.56& $^{8.63 }_{8.47}$ &    9.79  \\
       587739406242742472 &  178.020352 &   34.013853 &  0.3420 &  1095.0 &  \nodata&(\nodata) &   15.61&(0.91) &  \nodata& $^{... }_{...}$ &    8.86  \\
       587739828742389914 &  224.396405 &   22.533833 &  0.1488 &  1563.1 &    0.45&( 0.06) &    8.94&(0.29) &    8.55& $^{8.66 }_{8.41}$ &    9.30  \\
       587739652107600089 &  238.041673 &   21.053410 &  0.2332 &   893.5 &  \nodata&(\nodata) &    7.29&(0.44) &    8.68& $^{8.76 }_{8.57}$ &    9.90  \\
       587739721387409964 &  249.330431 &   14.651378 &  0.2923 &   452.5 &  \nodata&(\nodata) &    9.89&(0.17) &    8.62& $^{8.81 }_{8.33}$ &    9.64  \\
       587741600420003946 &  181.252807 &   26.346595 &  0.3427 &   305.2 &    1.06&( 0.24) &   18.45&(0.94) &    8.81& $^{8.88 }_{8.73}$ &    9.66  \\
       587741421099286852 &  126.715863 &   18.347732 &  0.2972 &   774.4 &    0.57&( 0.18) &    4.53&(0.20) &    8.75& $^{8.88 }_{8.58}$ &    8.67  \\
       587741532770074773 &  133.350354 &   19.506280 &  0.2365 &   860.3 &  \nodata&(\nodata) &    7.50&(0.41) &    8.68& $^{8.74 }_{8.61}$ &    9.37  \\
       587741817851084830 &  137.805603 &   18.518936 &  0.2622 &   329.8 &    1.72&( 0.16) &   26.17&(1.67) &    8.89& $^{8.92 }_{8.87}$ &    9.93  \\
       587741391573287017 &  145.946756 &   26.345161 &  0.2366 &    12.6 &  \nodata&(\nodata) &    3.15&(0.07) &    8.72& $^{8.79 }_{8.64}$ &   10.26  \\
       587741392649781464 &  152.329151 &   29.272638 &  0.2219 &  1042.8 &    0.34&( 0.09) &    4.51&(0.19) &    8.69& $^{8.80 }_{8.57}$ &    8.55  \\
       587739648351076573 &  155.239428 &   29.624017 &  0.2555 &   378.8 &  \nodata&(\nodata) &    6.17&(0.25) &    8.82& $^{8.85 }_{8.79}$ &    9.51  \\
       587741490367889543 &  158.112322 &   27.298680 &  0.1924 &   840.9 &    0.56&( 0.09) &   12.74&(0.41) &    8.69& $^{8.72 }_{8.66}$ &   10.02  \\
       587741532781215844 &  159.779860 &   27.472509 &  0.2801 &   182.2 &    0.81&( 0.11) &    4.57&(0.22) &  \nodata& $^{... }_{...}$ &    9.58  \\
       587742014876745993 &  141.869487 &   17.671838 &  0.2883 &   760.7 &    0.86&( 0.16) &   12.38&(0.36) &    8.51& $^{8.68 }_{8.26}$ &    9.38  \\
       588023240745943289 &  140.705287 &   19.227629 &  0.3175 &   526.8 &    0.45&( 0.13) &   14.38&(0.60) &    8.59& $^{8.76 }_{8.33}$ &   10.48  \\
       587745243087372534 &  141.384863 &   14.053623 &  0.3013 &  1430.4 &    0.97&( 0.13) &   18.73&(1.13) &    8.73& $^{8.83 }_{8.61}$ &    9.08  \\
       587742628534026489 &  243.276317 &    9.496990 &  0.2993 &   215.7 &  \nodata&(\nodata) &   14.14&(0.72) &    8.70& $^{8.80 }_{8.59}$ &    9.65  \\
       587744874785145599 &  121.325174 &    9.425978 &  0.3304 &   534.6 &  \nodata&(\nodata) &   21.67&(1.68) &  \nodata& $^{... }_{...}$ &   10.24  \\
       587742013825941802 &  197.653081 &   21.804731 &  0.2832 &   273.8 &    1.08&( 0.10) &    9.70&(0.68) &    8.83& $^{8.90 }_{8.75}$ &    9.27  \\
       587742062151467120 &  196.734804 &   22.694003 &  0.2741 &   879.9 &    0.38&( 0.12) &    8.83&(0.32) &  \nodata& $^{... }_{...}$ &   10.10  \\
       587741727655919734 &  193.761316 &   25.935911 &  0.3119 &   660.3 &    0.99&( 0.13) &    7.20&(0.43) &  \nodata& $^{... }_{...}$ &   10.02  \\
       \enddata
%\tablecomments{deluxetable}
\tablenotetext{a}{The values presented here are from the SDSS DR7 archive.}
\tablenotetext{b}{The values presented here are computed from the {\it GALEX} GR4 archive Far UV fluxes using the redshifts in Column 4.}
\tablenotetext{c}{The star formation rates presented here are from the H$\alpha$ line.  Errors are computed from 1 $\sigma$ flux errors. See \S~\ref{sec:sfr}. }
\tablenotetext{d}{The metallicities shown here are calculated from the [NII]/[OII] ratio \citep{kewleydopita2002}. The metallicity is followed by upper (above) and lower  (below) 1 $\sigma$ errors.  See \S~\ref{sec:met}.}
\tablenotetext{e}{The masses are calculated following the methodology of \citet{schawinskietal2007}. See \S~\ref{sec:mass}.}
\label{tab:sf}
\end{deluxetable}
\clearpage

\appendix

\bsp

\label{lastpage}


\begin{thebibliography}{113}
\expandafter\ifx\csname natexlab\endcsname\relax\def\natexlab#1{#1}\fi

\bibitem[{{Abazajian} {et~al.}(2009){Abazajian}, {Adelman-McCarthy},
  {Ag{\"u}eros}, {Allam}, {Allende Prieto}, {An}, {Anderson}, {Anderson},
  {Annis}, {Bahcall}, {Bailer-Jones}, {Barentine}, {Bassett}, {Becker},
  {Beers}, {Bell}, {Belokurov}, {Berlind}, {Berman}, {Bernardi}, {Bickerton},
  {Bizyaev}, {Blakeslee}, {Blanton}, {Bochanski}, {Boroski}, {Brewington},
  {Brinchmann}, {Brinkmann}, {Brunner}, {Budav{\'a}ri}, {Carey}, {Carliles},
  {Carr}, {Castander}, {Cinabro}, {Connolly}, {Csabai}, {Cunha}, {Czarapata},
  {Davenport}, {de Haas}, {Dilday}, {Doi}, {Eisenstein}, {Evans}, {Evans},
  {Fan}, {Friedman}, {Frieman}, {Fukugita}, {G{\"a}nsicke}, {Gates},
  {Gillespie}, {Gilmore}, {Gonzalez}, {Gonzalez}, {Grebel}, {Gunn},
  {Gy{\"o}ry}, {Hall}, {Harding}, {Harris}, {Harvanek}, {Hawley}, {Hayes},
  {Heckman}, {Hendry}, {Hennessy}, {Hindsley}, {Hoblitt}, {Hogan}, {Hogg},
  {Holtzman}, {Hyde}, {Ichikawa}, {Ichikawa}, {Im}, {Ivezi{\'c}}, {Jester},
  {Jiang}, {Johnson}, {Jorgensen}, {Juri{\'c}}, {Kent}, {Kessler}, {Kleinman},
  {Knapp}, {Konishi}, {Kron}, {Krzesinski}, {Kuropatkin}, {Lampeitl},
  {Lebedeva}, {Lee}, {Lee}, {Leger}, {L{\'e}pine}, {Li}, {Lima}, {Lin}, {Long},
  {Loomis}, {Loveday}, {Lupton}, {Magnier}, {Malanushenko}, {Malanushenko},
  {Mandelbaum}, {Margon}, {Marriner}, {Mart{\'{\i}}nez-Delgado}, {Matsubara},
  {McGehee}, {McKay}, {Meiksin}, {Morrison}, {Mullally}, {Munn}, {Murphy},
  {Nash}, {Nebot}, {Neilsen}, {Newberg}, {Newman}, {Nichol}, {Nicinski},
  {Nieto-Santisteban}, {Nitta}, {Okamura}, {Oravetz}, {Ostriker}, {Owen},
  {Padmanabhan}, {Pan}, {Park}, {Pauls}, {Peoples}, {Percival}, {Pier}, {Pope},
  {Pourbaix}, {Price}, {Purger}, {Quinn}, {Raddick}, {Fiorentin}, {Richards},
  {Richmond}, {Riess}, {Rix}, {Rockosi}, {Sako}, {Schlegel}, {Schneider},
  {Scholz}, {Schreiber}, {Schwope}, {Seljak}, {Sesar}, {Sheldon}, {Shimasaku},
  {Sibley}, {Simmons}, {Sivarani}, {Smith}, {Smith}, {Smol{\v c}i{\'c}},
  {Snedden}, {Stebbins}, {Steinmetz}, {Stoughton}, {Strauss}, {Subba Rao},
  {Suto}, {Szalay}, {Szapudi}, {Szkody}, {Tanaka}, {Tegmark}, {Teodoro},
  {Thakar}, {Tremonti}, {Tucker}, {Uomoto}, {Vanden Berk}, {Vandenberg},
  {Vidrih}, {Vogeley}, {Voges}, {Vogt}, {Wadadekar}, {Watters}, {Weinberg},
  {West}, {White}, {Wilhite}, {Wonders}, {Yanny}, {Yocum}, {York}, {Zehavi},
  {Zibetti}, \& {Zucker}}]{abazajianetal2009}
{Abazajian} K.~N., {Adelman-McCarthy} J.~K., {Ag{\"u}eros} M.~A., {Allam}
  S.~S., {Allende Prieto} C., {An} D., {Anderson} K.~S.~J., {Anderson} S.~F.,
  {Annis} J., {Bahcall} N.~A., {Bailer-Jones} C.~A.~L., {Barentine} J.~C.,
  {Bassett} B.~A., {Becker} A.~C., {Beers} T.~C., {Bell} E.~F., {Belokurov} V.,
  {Berlind} A.~A., {Berman} E.~F., {Bernardi} M., {Bickerton} S.~J., {Bizyaev}
  D., {Blakeslee} J.~P., {Blanton} M.~R., {Bochanski} J.~J., {Boroski} W.~N.,
  {Brewington} H.~J., {Brinchmann} J., {Brinkmann} J., {Brunner} R.~J.,
  {Budav{\'a}ri} T., {Carey} L.~N., {Carliles} S., {Carr} M.~A., {Castander}
  F.~J., {Cinabro} D., {Connolly} A.~J., {Csabai} I., {Cunha} C.~E.,
  {Czarapata} P.~C., {Davenport} J.~R.~A., {de Haas} E., {Dilday} B., {Doi} M.,
  {Eisenstein} D.~J., {Evans} M.~L., {Evans} N.~W., {Fan} X., {Friedman} S.~D.,
  {Frieman} J.~A., {Fukugita} M., {G{\"a}nsicke} B.~T., {Gates} E., {Gillespie}
  B., {Gilmore} G., {Gonzalez} B., {Gonzalez} C.~F., {Grebel} E.~K., {Gunn}
  J.~E., {Gy{\"o}ry} Z., {Hall} P.~B., {Harding} P., {Harris} F.~H., {Harvanek}
  M., {Hawley} S.~L., {Hayes} J.~J.~E., {Heckman} T.~M., {Hendry} J.~S.,
  {Hennessy} G.~S., {Hindsley} R.~B., {Hoblitt} J., {Hogan} C.~J., {Hogg}
  D.~W., {Holtzman} J.~A., {Hyde} J.~B., {Ichikawa} S.-i., {Ichikawa} T., {Im}
  M., {Ivezi{\'c}} {\v Z}., {Jester} S., {Jiang} L., {Johnson} J.~A.,
  {Jorgensen} A.~M., {Juri{\'c}} M., {Kent} S.~M., {Kessler} R., {Kleinman}
  S.~J., {Knapp} G.~R., {Konishi} K., {Kron} R.~G., {Krzesinski} J.,
  {Kuropatkin} N., {Lampeitl} H., {Lebedeva} S., {Lee} M.~G., {Lee} Y.~S.,
  {Leger} R.~F., {L{\'e}pine} S., {Li} N., {Lima} M., {Lin} H., {Long} D.~C.,
  {Loomis} C.~P., {Loveday} J., {Lupton} R.~H., {Magnier} E., {Malanushenko}
  O., {Malanushenko} V., {Mandelbaum} R., {Margon} B., {Marriner} J.~P.,
  {Mart{\'{\i}}nez-Delgado} D., {Matsubara} T., {McGehee} P.~M., {McKay} T.~A.,
  {Meiksin} A., {Morrison} H.~L., {Mullally} F., {Munn} J.~A., {Murphy} T.,
  {Nash} T., {Nebot} A., {Neilsen} E.~H., {Newberg} H.~J., {Newman} P.~R.,
  {Nichol} R.~C., {Nicinski} T., {Nieto-Santisteban} M., {Nitta} A., {Okamura}
  S., {Oravetz} D.~J., {Ostriker} J.~P., {Owen} R., {Padmanabhan} N., {Pan} K.,
  {Park} C., {Pauls} G., {Peoples} J., {Percival} W.~J., {Pier} J.~R., {Pope}
  A.~C., {Pourbaix} D., {Price} P.~A., {Purger} N., {Quinn} T., {Raddick}
  M.~J., {Fiorentin} P.~R., {Richards} G.~T., {Richmond} M.~W., {Riess} A.~G.,
  {Rix} H.-W., {Rockosi} C.~M., {Sako} M., {Schlegel} D.~J., {Schneider} D.~P.,
  {Scholz} R.-D., {Schreiber} M.~R., {Schwope} A.~D., {Seljak} U., {Sesar} B.,
  {Sheldon} E., {Shimasaku} K., {Sibley} V.~C., {Simmons} A.~E., {Sivarani} T.,
  {Smith} J.~A., {Smith} M.~C., {Smol{\v c}i{\'c}} V., {Snedden} S.~A.,
  {Stebbins} A., {Steinmetz} M., {Stoughton} C., {Strauss} M.~A., {Subba Rao}
  M., {Suto} Y., {Szalay} A.~S., {Szapudi} I., {Szkody} P., {Tanaka} M.,
  {Tegmark} M., {Teodoro} L.~F.~A., {Thakar} A.~R., {Tremonti} C.~A., {Tucker}
  D.~L., {Uomoto} A., {Vanden Berk} D.~E., {Vandenberg} J., {Vidrih} S.,
  {Vogeley} M.~S., {Voges} W., {Vogt} N.~P., {Wadadekar} Y., {Watters} S.,
  {Weinberg} D.~H., {West} A.~A., {White} S.~D.~M., {Wilhite} B.~C., {Wonders}
  A.~C., {Yanny} B., {Yocum} D.~R., {York} D.~G., {Zehavi} I., {Zibetti} S.,
  {Zucker} D.~B., 2009, \apjs, 182, 543

\bibitem[{{Aloisi} {et~al.}(2007){Aloisi}, {Clementini}, {Tosi}, {Annibali},
  {Contreras}, {Fiorentino}, {Mack}, {Marconi}, {Musella}, {Saha}, {Sirianni},
  \& {van der Marel}}]{aloisietal2007}
{Aloisi} A., {Clementini} G., {Tosi} M., {Annibali} F., {Contreras} R.,
  {Fiorentino} G., {Mack} J., {Marconi} M., {Musella} I., {Saha} A., {Sirianni}
  M., {van der Marel} R.~P., 2007, \apjl, 667, L151

\bibitem[{{Asplund} {et~al.}(2005){Asplund}, {Grevesse}, \&
  {Sauval}}]{asplundetal2005}
{Asplund} M., {Grevesse} N., {Sauval} A.~J., 2005, in Astronomical Society of
  the Pacific Conference Series, Vol. 336, Cosmic Abundances as Records of
  Stellar Evolution and Nucleosynthesis, {Barnes} III T.~G., {Bash} F.~N.,
  eds., pp. 25--+

\bibitem[{{Bamford} {et~al.}(2009){Bamford}, {Nichol}, {Baldry}, {Land},
  {Lintott}, {Schawinski}, {Slosar}, {Szalay}, {Thomas}, {Torki}, {Andreescu},
  {Edmondson}, {Miller}, {Murray}, {Raddick}, \&
  {Vandenberg}}]{bamfordetal2009}
{Bamford} S.~P., {Nichol} R.~C., {Baldry} I.~K., {Land} K., {Lintott} C.~J.,
  {Schawinski} K., {Slosar} A., {Szalay} A.~S., {Thomas} D., {Torki} M.,
  {Andreescu} D., {Edmondson} E.~M., {Miller} C.~J., {Murray} P., {Raddick}
  M.~J., {Vandenberg} J., 2009, \mnras, 393, 1324

\bibitem[{{Barmby} {et~al.}(2004){Barmby}, {Huang}, {Fazio}, {Surace},
  {Arendt}, {Hora}, {Pahre}, {Adelberger}, {Eisenhardt}, {Erb}, {Pettini},
  {Reach}, {Reddy}, {Shapley}, {Steidel}, {Stern}, {Wang}, \&
  {Willner}}]{barmbyetal2004}
{Barmby} P., {Huang} J.-S., {Fazio} G.~G., {Surace} J.~A., {Arendt} R.~G.,
  {Hora} J.~L., {Pahre} M.~A., {Adelberger} K.~L., {Eisenhardt} P., {Erb}
  D.~K., {Pettini} M., {Reach} W.~T., {Reddy} N.~A., {Shapley} A.~E., {Steidel}
  C.~C., {Stern} D., {Wang} Z., {Willner} S.~P., 2004, \apjs, 154, 97

\bibitem[{{Basu} \& {Antia}(2008)}]{basuantia2008}
{Basu} S., {Antia} H.~M., 2008, Physics Reports, 457, 217

\bibitem[{{Bauer} {et~al.}(2005){Bauer}, {Drory}, {Hill}, \&
  {Feulner}}]{baueretal2005}
{Bauer} A.~E., {Drory} N., {Hill} G.~J., {Feulner} G., 2005, \apjl, 621, L89

\bibitem[{{Bergvall} \& {Johansson}(1985)}]{bergvalljohansson1985}
{Bergvall} N., {Johansson} L., 1985, \aap, 149, 475

\bibitem[{{Bergvall} \& {{\"O}stlin}(2002)}]{bergvallostlin2002}
{Bergvall} N., {{\"O}stlin} G., 2002, \aap, 390, 891

\bibitem[{{Blanton} {et~al.}(2001){Blanton}, {Dalcanton}, {Eisenstein},
  {Loveday}, {Strauss}, {SubbaRao}, {Weinberg}, {Anderson}, {Annis}, {Bahcall},
  \& et~al.}]{blantonetal2001}
{Blanton} M.~R., {Dalcanton} J., {Eisenstein} D., {Loveday} J., {Strauss}
  M.~A., {SubbaRao} M., {Weinberg} D.~H., {Anderson} Jr. J.~E., {Annis} J.,
  {Bahcall} N.~A., et~al., 2001, \aj, 121, 2358

\bibitem[{{Bonning} {et~al.}(2005){Bonning}, {Shields}, {Salviander}, \&
  {McLure}}]{bonningetal2005}
{Bonning} E.~W., {Shields} G.~A., {Salviander} S., {McLure} R.~J., 2005, \apj,
  626, 89

\bibitem[{{Bremer} {et~al.}(2004){Bremer}, {Lehnert}, {Waddington},
  {Hardcastle}, {Boyce}, \& {Phillipps}}]{bremeretal2004}
{Bremer} M.~N., {Lehnert} M.~D., {Waddington} I., {Hardcastle} M.~J., {Boyce}
  P.~J., {Phillipps} S., 2004, \mnras, 347, L7

\bibitem[{{Brinchmann} {et~al.}(2004){Brinchmann}, {Charlot}, {White},
  {Tremonti}, {Kauffmann}, {Heckman}, \& {Brinkmann}}]{brinkmannetal2004}
{Brinchmann} J., {Charlot} S., {White} S.~D.~M., {Tremonti} C., {Kauffmann} G.,
  {Heckman} T., {Brinkmann} J., 2004, \mnras, 351, 1151

\bibitem[{{Calzetti} {et~al.}(2000){Calzetti}, {Armus}, {Bohlin}, {Kinney},
  {Koornneef}, \& {Storchi-Bergmann}}]{calzettietal2000}
{Calzetti} D., {Armus} L., {Bohlin} R.~C., {Kinney} A.~L., {Koornneef} J.,
  {Storchi-Bergmann} T., 2000, \apj, 533, 682

\bibitem[{{Cappellari} \& {Emsellem}(2004)}]{cappellariemsellem2004}
{Cappellari} M., {Emsellem} E., 2004, \pasp, 116, 138

\bibitem[{{Cardelli} {et~al.}(1989){Cardelli}, {Clayton}, \&
  {Mathis}}]{cardelli1989}
{Cardelli} J.~A., {Clayton} G.~C., {Mathis} J.~S., 1989, \apj, 345, 245

\bibitem[{{Carilli} {et~al.}(2008){Carilli}, {Lee}, {Capak}, {Schinnerer},
  {Lee}, {McCraken}, {Yun}, {Scoville}, {Smol{\v c}i{\'c}}, {Giavalisco},
  {Datta}, {Taniguchi}, \& {Urry}}]{carillietal2008}
{Carilli} C.~L., {Lee} N., {Capak} P., {Schinnerer} E., {Lee} K.-S., {McCraken}
  H., {Yun} M.~S., {Scoville} N., {Smol{\v c}i{\'c}} V., {Giavalisco} M.,
  {Datta} A., {Taniguchi} Y., {Urry} C.~M., 2008, \apj, 689, 883

\bibitem[{{Castro Cer{\'o}n} {et~al.}(2006){Castro Cer{\'o}n},
  {Micha{\l}owski}, {Hjorth}, {Watson}, {Fynbo}, \&
  {Gorosabel}}]{castroceronetal2006}
{Castro Cer{\'o}n} J.~M., {Micha{\l}owski} M.~J., {Hjorth} J., {Watson} D.,
  {Fynbo} J.~P.~U., {Gorosabel} J., 2006, \apjl, 653, L85

\bibitem[{{Coppin} {et~al.}(2007){Coppin}, {Swinbank}, {Neri}, {Cox}, {Smail},
  {Ellis}, {Geach}, {Siana}, {Teplitz}, {Dye}, {Kneib}, {Edge}, \&
  {Richard}}]{coppinetal2007}
{Coppin} K.~E.~K., {Swinbank} A.~M., {Neri} R., {Cox} P., {Smail} I., {Ellis}
  R.~S., {Geach} J.~E., {Siana} B., {Teplitz} H., {Dye} S., {Kneib} J.-P.,
  {Edge} A.~C., {Richard} J., 2007, \apj, 665, 936

\bibitem[{{Corbin} \& {Vacca}(2002)}]{corbinvacca2002}
{Corbin} M.~R., {Vacca} W.~D., 2002, \apj, 581, 1039

\bibitem[{{Corbin} {et~al.}(2006){Corbin}, {Vacca}, {Cid Fernandes}, {Hibbard},
  {Somerville}, \& {Windhorst}}]{corbinetal2006}
{Corbin} M.~R., {Vacca} W.~D., {Cid Fernandes} R., {Hibbard} J.~E.,
  {Somerville} R.~S., {Windhorst} R.~A., 2006, \apj, 651, 861

\bibitem[{{Cowie} \& {Hu}(1998)}]{cowieandhu1998}
{Cowie} L.~L., {Hu} E.~M., 1998, \aj, 115, 1319

\bibitem[{{Cowie} {et~al.}(1996){Cowie}, {Songaila}, {Hu}, \&
  {Cohen}}]{cowieetal1996}
{Cowie} L.~L., {Songaila} A., {Hu} E.~M., {Cohen} J.~G., 1996, \aj, 112, 839

\bibitem[{{Darg} {et~al.}(2009){Darg}, {Kaviraj}, {Lintott}, {Schawinski},
  {Sarzi}, {Bamford}, {Silk}, {Andreescu}, {Murray}, {Nichol}, {Raddick},
  {Slosar}, {Szalay}, {Thomas}, \& {Vandenberg}}]{dargetal2009a}
{Darg} D.~W., {Kaviraj} S., {Lintott} C.~J., {Schawinski} K., {Sarzi} M.,
  {Bamford} S., {Silk} J., {Andreescu} D., {Murray} P., {Nichol} R.~C.,
  {Raddick} M.~J., {Slosar} A., {Szalay} A.~S., {Thomas} D., {Vandenberg} J.,
  2009, ArXiv:0903.5057

\bibitem[{{Finkelstein} {et~al.}(2007){Finkelstein}, {Rhoads}, {Malhotra},
  {Pirzkal}, \& {Wang}}]{finkelsteinetal2007}
{Finkelstein} S.~L., {Rhoads} J.~E., {Malhotra} S., {Pirzkal} N., {Wang} J.,
  2007, \apj, 660, 1023

\bibitem[{{Fukugita} {et~al.}(1996){Fukugita}, {Ichikawa}, {Gunn}, {Doi},
  {Shimasaku}, \& {Schneider}}]{fukugitaetal1996}
{Fukugita} M., {Ichikawa} T., {Gunn} J.~E., {Doi} M., {Shimasaku} K.,
  {Schneider} D.~P., 1996, \aj, 111, 1748

\bibitem[{{Gawiser} {et~al.}(2007){Gawiser}, {Francke}, {Lai}, {Schawinski},
  {Gronwall}, {Ciardullo}, {Quadri}, {Orsi}, {Barrientos}, {Blanc}, {Fazio},
  {Feldmeier}, {Huang}, {Infante}, {Lira}, {Padilla}, {Taylor}, {Treister},
  {Urry}, {van Dokkum}, \& {Virani}}]{gawiseretal2007}
{Gawiser} E., {Francke} H., {Lai} K., {Schawinski} K., {Gronwall} C.,
  {Ciardullo} R., {Quadri} R., {Orsi} A., {Barrientos} L.~F., {Blanc} G.~A.,
  {Fazio} G., {Feldmeier} J.~J., {Huang} J.-s., {Infante} L., {Lira} P.,
  {Padilla} N., {Taylor} E.~N., {Treister} E., {Urry} C.~M., {van Dokkum}
  P.~G., {Virani} S.~N., 2007, \apj, 671, 278

\bibitem[{{Gawiser} {et~al.}(2006){Gawiser}, {van Dokkum}, {Gronwall},
  {Ciardullo}, {Blanc}, {Castander}, {Feldmeier}, {Francke}, {Franx},
  {Haberzettl}, {Herrera}, {Hickey}, {Infante}, {Lira}, {Maza}, {Quadri},
  {Richardson}, {Schawinski}, {Schirmer}, {Taylor}, {Treister}, {Urry}, \&
  {Virani}}]{gawiseretal2006lya}
{Gawiser} E., {van Dokkum} P.~G., {Gronwall} C., {Ciardullo} R., {Blanc} G.~A.,
  {Castander} F.~J., {Feldmeier} J., {Francke} H., {Franx} M., {Haberzettl} L.,
  {Herrera} D., {Hickey} T., {Infante} L., {Lira} P., {Maza} J., {Quadri} R.,
  {Richardson} A., {Schawinski} K., {Schirmer} M., {Taylor} E.~N., {Treister}
  E., {Urry} C.~M., {Virani} S.~N., 2006, \apjl, 642, L13

\bibitem[{{Giavalisco}(2002)}]{giavaliscoetal2002}
{Giavalisco} M., 2002, \araa, 40, 579

\bibitem[{{Giavalisco} {et~al.}(2004){Giavalisco}, {Dickinson}, {Ferguson},
  {Ravindranath}, {Kretchmer}, {Moustakas}, {Madau}, {Fall}, {Gardner},
  {Livio}, {Papovich}, {Renzini}, {Spinrad}, {Stern}, \&
  {Riess}}]{giavaliscoetal2004}
{Giavalisco} M., {Dickinson} M., {Ferguson} H.~C., {Ravindranath} S.,
  {Kretchmer} C., {Moustakas} L.~A., {Madau} P., {Fall} S.~M., {Gardner} J.~P.,
  {Livio} M., {Papovich} C., {Renzini} A., {Spinrad} H., {Stern} D., {Riess}
  A., 2004, \apjl, 600, L103

\bibitem[{{Giavalisco} {et~al.}(1996){Giavalisco}, {Livio}, {Bohlin},
  {Macchetto}, \& {Stecher}}]{giavaliscoetal1996}
{Giavalisco} M., {Livio} M., {Bohlin} R.~C., {Macchetto} F.~D., {Stecher}
  T.~P., 1996, \aj, 112, 369

\bibitem[{{Gil de Paz} \& {Madore}(2005)}]{gildepaz2005}
{Gil de Paz} A., {Madore} B.~F., 2005, \apjs, 156, 345

\bibitem[{{Grevesse} \& {Sauval}(1998)}]{grevesseetal1998}
{Grevesse} N., {Sauval} A.~J., 1998, Space Science Reviews, 85, 161

\bibitem[{{Gronwall} {et~al.}(2007){Gronwall}, {Ciardullo}, {Hickey},
  {Gawiser}, {Feldmeier}, {van Dokkum}, {Urry}, {Herrera}, {Lehmer}, {Infante},
  {Orsi}, {Marchesini}, {Blanc}, {Francke}, {Lira}, \&
  {Treister}}]{gronwalletal2007}
{Gronwall} C., {Ciardullo} R., {Hickey} T., {Gawiser} E., {Feldmeier} J.~J.,
  {van Dokkum} P.~G., {Urry} C.~M., {Herrera} D., {Lehmer} B.~D., {Infante} L.,
  {Orsi} A., {Marchesini} D., {Blanc} G.~A., {Francke} H., {Lira} P.,
  {Treister} E., 2007, \apj, 667, 79

\bibitem[{{Guseva} {et~al.}(2004){Guseva}, {Papaderos}, {Izotov}, {Noeske}, \&
  {Fricke}}]{gusevaetal2004}
{Guseva} N.~G., {Papaderos} P., {Izotov} Y.~I., {Noeske} K.~G., {Fricke} K.~J.,
  2004, \aap, 421, 519

\bibitem[{{Guzman} {et~al.}(1997){Guzman}, {Gallego}, {Koo}, {Phillips},
  {Lowenthal}, {Faber}, {Illingworth}, \& {Vogt}}]{guzmanetal1997}
{Guzman} R., {Gallego} J., {Koo} D.~C., {Phillips} A.~C., {Lowenthal} J.~D.,
  {Faber} S.~M., {Illingworth} G.~D., {Vogt} N.~P., 1997, \apj, 489, 559

\bibitem[{{Guzm{\'a}n} {et~al.}(2003){Guzm{\'a}n}, {{\"O}stlin}, {Kunth},
  {Bershady}, {Koo}, \& {Pahre}}]{guzmanetal2003}
{Guzm{\'a}n} R., {{\"O}stlin} G., {Kunth} D., {Bershady} M.~A., {Koo} D.~C.,
  {Pahre} M.~A., 2003, \apjl, 586, L45

\bibitem[{{Heckman} {et~al.}(2005){Heckman}, {Hoopes}, {Seibert}, {Martin},
  {Salim}, {Rich}, {Kauffmann}, {Charlot}, {Barlow}, {Bianchi}, {Byun}, \&
  {Donas}}]{heckmanetal2005}
{Heckman} T.~M., {Hoopes} C.~G., {Seibert} M., {Martin} D.~C., {Salim} S.,
  {Rich} R.~M., {Kauffmann} G., {Charlot} S., {Barlow} T.~A., {Bianchi} L.,
  {Byun} Y.-I., {Donas} J., 2005, \apjl, 619, L35

\bibitem[{{Hoopes} {et~al.}(2007){Hoopes}, {Heckman}, {Salim}, {Seibert},
  {Tremonti}, {Schiminovich}, {Rich}, {Lee}, {Madore}, {Milliard}, {Szalay},
  {Welsh}, \& {Yi}}]{hoopesetal2007}
{Hoopes} C.~G., {Heckman} T.~M., {Salim} S., {Seibert} M., {Tremonti} C.~A.,
  {Schiminovich} D., {Rich} R.~M., {Lee} Y.-W., {Madore} B.~F., {Milliard} B.,
  {Szalay} A.~S., {Welsh} B.~Y., {Yi} S.~K., 2007, \apjs, 173, 441

\bibitem[{{Hornschemeier} {et~al.}(2008){Hornschemeier}, {Heckman}, {Ptak},
  {Grimes}, {Strickland}, {Salim}, {Rich}, \&
  {Mallery}}]{Hornschemeieretal2008}
{Hornschemeier} A.~E., {Heckman} T., {Ptak} A., {Grimes} J., {Strickland} D.,
  {Salim} S., {Rich} R.~M., {Mallery} R., 2008, in American Institute of
  Physics Conference Series, Vol. 1010, A Population Explosion: The Nature \&
  Evolution of X-ray Binaries in Diverse Environments, {Bandyopadhyay} R.~M.,
  {Wachter} S., {Gelino} D., {Gelino} C.~R., eds., pp. 291--297

\bibitem[{{Hoyos} {et~al.}(2004){Hoyos}, {Guzm{\'a}n}, {Bershady}, {Koo}, \&
  {D{\'{\i}}az}}]{hoyosetal2004}
{Hoyos} C., {Guzm{\'a}n} R., {Bershady} M.~A., {Koo} D.~C., {D{\'{\i}}az}
  A.~I., 2004, \aj, 128, 1541

\bibitem[{{Hoyos} {et~al.}(2007){Hoyos}, {Guzm{\'a}n}, {D{\'{\i}}az}, {Koo}, \&
  {Bershady}}]{hoyosetal2007}
{Hoyos} C., {Guzm{\'a}n} R., {D{\'{\i}}az} A.~I., {Koo} D.~C., {Bershady}
  M.~A., 2007, \aj, 134, 2455

\bibitem[{{Izotov} {et~al.}(2006){Izotov}, {Stasi{\'n}ska}, {Meynet}, {Guseva},
  \& {Thuan}}]{izotovetal2006}
{Izotov} Y.~I., {Stasi{\'n}ska} G., {Meynet} G., {Guseva} N.~G., {Thuan} T.~X.,
  2006, \aap, 448, 955

\bibitem[{{Kakazu} {et~al.}(2007){Kakazu}, {Cowie}, \& {Hu}}]{kakazuetal2007}
{Kakazu} Y., {Cowie} L.~L., {Hu} E.~M., 2007, \apj, 668, 853

\bibitem[{{Kauffmann} {et~al.}(2003){Kauffmann}, {Heckman}, {Tremonti},
  {Brinchmann}, {Charlot}, {White}, {Ridgway}, {Brinkmann}, {Fukugita}, {Hall},
  {Ivezi{\'c}}, {Richards}, \& {Schneider}}]{kauffmannetal2003}
{Kauffmann} G., {Heckman} T.~M., {Tremonti} C., {Brinchmann} J., {Charlot} S.,
  {White} S.~D.~M., {Ridgway} S.~E., {Brinkmann} J., {Fukugita} M., {Hall}
  P.~B., {Ivezi{\'c}} {\v Z}., {Richards} G.~T., {Schneider} D.~P., 2003,
  \mnras, 346, 1055

\bibitem[{{Kennicutt}(1998)}]{kennicutt1998}
{Kennicutt} Jr. R.~C., 1998, \araa, 36, 189

\bibitem[{{Kewley} \& {Dopita}(2002)}]{kewleydopita2002}
{Kewley} L.~J., {Dopita} M.~A., 2002, \apjs, 142, 35

\bibitem[{{Kewley} {et~al.}(2001){Kewley}, {Dopita}, {Sutherland}, {Heisler},
  \& {Trevena}}]{kewleyetal2001}
{Kewley} L.~J., {Dopita} M.~A., {Sutherland} R.~S., {Heisler} C.~A., {Trevena}
  J., 2001, \apj, 556, 121

\bibitem[{{Kewley} {et~al.}(2006){Kewley}, {Groves}, {Kauffmann}, \&
  {Heckman}}]{kewleyetal2006}
{Kewley} L.~J., {Groves} B., {Kauffmann} G., {Heckman} T., 2006, \mnras, 372,
  961

\bibitem[{{Komossa} {et~al.}(2008){Komossa}, {Xu}, {Zhou}, {Storchi-Bergmann},
  \& {Binette}}]{komossaetal2008}
{Komossa} S., {Xu} D., {Zhou} H., {Storchi-Bergmann} T., {Binette} L., 2008,
  \apj, 680, 926

\bibitem[{{Kriek} {et~al.}(2009){Kriek}, {van Dokkum}, {Labbe}, {Franx},
  {Illingworth}, {Marchesini}, \& {Quadri}}]{krieketal2009}
{Kriek} M., {van Dokkum} P.~G., {Labbe} I., {Franx} M., {Illingworth} G.~D.,
  {Marchesini} D., {Quadri} R.~F., 2009, ArXiv e-prints

\bibitem[{{Kudritzki} {et~al.}(2000){Kudritzki}, {M{\'e}ndez}, {Feldmeier},
  {Ciardullo}, {Jacoby}, {Freeman}, {Arnaboldi}, {Capaccioli}, {Gerhard}, \&
  {Ford}}]{kudritzkietal2000}
{Kudritzki} R.-P., {M{\'e}ndez} R.~H., {Feldmeier} J.~J., {Ciardullo} R.,
  {Jacoby} G.~H., {Freeman} K.~C., {Arnaboldi} M., {Capaccioli} M., {Gerhard}
  O., {Ford} H.~C., 2000, \apj, 536, 19

\bibitem[{{Kunth} {et~al.}(1988){Kunth}, {Maurogordato}, \&
  {Vigroux}}]{kunthetal1988}
{Kunth} D., {Maurogordato} S., {Vigroux} L., 1988, \aap, 204, 10

\bibitem[{{Land} {et~al.}(2008){Land}, {Slosar}, {Lintott}, {Andreescu},
  {Bamford}, {Murray}, {Nichol}, {Raddick}, {Schawinski}, {Szalay}, {Thomas},
  \& {Vandenberg}}]{landetal2008}
{Land} K., {Slosar} A., {Lintott} C., {Andreescu} D., {Bamford} S., {Murray}
  P., {Nichol} R., {Raddick} M.~J., {Schawinski} K., {Szalay} A., {Thomas} D.,
  {Vandenberg} J., 2008, \mnras, 388, 1686

\bibitem[{{Lehmer} {et~al.}(2005){Lehmer}, {Brandt}, {Alexander}, {Bauer},
  {Conselice}, {Dickinson}, {Giavalisco}, {Grogin}, {Koekemoer}, {Lee},
  {Moustakas}, \& {Schneider}}]{lehmeretal2005lbg}
{Lehmer} B.~D., {Brandt} W.~N., {Alexander} D.~M., {Bauer} F.~E., {Conselice}
  C.~J., {Dickinson} M.~E., {Giavalisco} M., {Grogin} N.~A., {Koekemoer} A.~M.,
  {Lee} K.-S., {Moustakas} L.~A., {Schneider} D.~P., 2005, \aj, 129, 1

\bibitem[{{Lequeux}(2005)}]{lequeux2005}
{Lequeux} J., 2005, {The interstellar medium}. The interstellar medium,
  Translation from the French language edition of: Le Milieu Interstellaire by
  James Lequeux, EDP Sciences, 2003 Edited by J.~Lequeux.~ Astronomy and
  astrophysics library, Berlin: Springer, 2005

\bibitem[{{Lilly} {et~al.}(1996){Lilly}, {Le Fevre}, {Hammer}, \&
  {Crampton}}]{lillyetal1996}
{Lilly} S.~J., {Le Fevre} O., {Hammer} F., {Crampton} D., 1996, \apjl, 460, L1+

\bibitem[{{Lintott} {et~al.}(2008){Lintott}, {Schawinski}, {Slosar}, {Land},
  {Bamford}, {Thomas}, {Raddick}, {Nichol}, {Szalay}, {Andreescu}, {Murray}, \&
  {Vandenberg}}]{lintottetal2008}
{Lintott} C.~J., {Schawinski} K., {Slosar} A., {Land} K., {Bamford} S.,
  {Thomas} D., {Raddick} M.~J., {Nichol} R.~C., {Szalay} A., {Andreescu} D.,
  {Murray} P., {Vandenberg} J., 2008, \mnras, 389, 1179

\bibitem[{{Lintott}(2009)}]{lintottetal2009}
{Lintott} C.~J. e.~a., 2009, in prep,

\bibitem[{{Loose} \& {Thuan}(1986)}]{looseetal1986}
{Loose} H.-H., {Thuan} T.~X., 1986, \apj, 309, 59

\bibitem[{{Lotz} {et~al.}(2006){Lotz}, {Madau}, {Giavalisco}, {Primack}, \&
  {Ferguson}}]{lotzetal2006}
{Lotz} J.~M., {Madau} P., {Giavalisco} M., {Primack} J., {Ferguson} H.~C.,
  2006, \apj, 636, 592

\bibitem[{{Lupton} {et~al.}(2001){Lupton}, {Gunn}, {Ivezi{\'c}}, {Knapp}, \&
  {Kent}}]{luptonetal2001}
{Lupton} R., {Gunn} J.~E., {Ivezi{\'c}} Z., {Knapp} G.~R., {Kent} S., 2001, in
  Astronomical Society of the Pacific Conference Series, Vol. 238, Astronomical
  Data Analysis Software and Systems X, {Harnden} Jr. F.~R., {Primini} F.~A.,
  {Payne} H.~E., eds., pp. 269--+

\bibitem[{{Madau} {et~al.}(1996){Madau}, {Ferguson}, {Dickinson}, {Giavalisco},
  {Steidel}, \& {Fruchter}}]{madauetal1996}
{Madau} P., {Ferguson} H.~C., {Dickinson} M.~E., {Giavalisco} M., {Steidel}
  C.~C., {Fruchter} A., 1996, \mnras, 283, 1388

\bibitem[{{Mannucci} {et~al.}(2009){Mannucci}, {Cresci}, {Maiolino}, {Marconi},
  {Pastorini}, {Pozzetti}, {Gnerucci}, {Risaliti}, {Schneider}, {Lehnert}, \&
  {Salvati}}]{mannuccietal2009}
{Mannucci} F., {Cresci} G., {Maiolino} R., {Marconi} A., {Pastorini} G.,
  {Pozzetti} L., {Gnerucci} A., {Risaliti} G., {Schneider} R., {Lehnert} M.,
  {Salvati} M., 2009, ArXiv:0902.2398

\bibitem[{{Maraston}(1998)}]{marastonetal1998}
{Maraston} C., 1998, \mnras, 300, 872

\bibitem[{{Maraston}(2005)}]{maraston2005}
---, 2005, \mnras, 362, 799

\bibitem[{{Marconi} {et~al.}(2008){Marconi}, {Axon}, {Maiolino}, {Nagao},
  {Pastorini}, {Pietrini}, {Robinson}, \& {Torricelli}}]{marconietal2008}
{Marconi} A., {Axon} D.~J., {Maiolino} R., {Nagao} T., {Pastorini} G.,
  {Pietrini} P., {Robinson} A., {Torricelli} G., 2008, \apj, 678, 693

\bibitem[{{McLure} \& {Jarvis}(2002)}]{mclurejarvis2002}
{McLure} R.~J., {Jarvis} M.~J., 2002, \mnras, 337, 109

\bibitem[{{Morrissey} {et~al.}(2007){Morrissey}, {Conrow}, {Barlow}, {Small},
  {Seibert}, {Wyder}, {Budav{\'a}ri}, {Rich}, {Szalay}, {Welsh}, \&
  {Yi}}]{morrisseyetal2007}
{Morrissey} P., {Conrow} T., {Barlow} T.~A., {Small} T., {Seibert} M., {Wyder}
  T.~K., {Budav{\'a}ri} T., {Rich} R.~M., {Szalay} A.~S., {Welsh} B.~Y., {Yi}
  S.~K., 2007, \apjs, 173, 682

\bibitem[{{Noll}(2007)}]{noll2007}
{Noll} K., 2007, in HST Proposal, pp. 11113--+

\bibitem[{{Osterbrock} \& {Pogge}(1985)}]{osterbrockandpogge1985}
{Osterbrock} D.~E., {Pogge} R.~W., 1985, \apj, 297, 166

\bibitem[{{Ouchi} {et~al.}(2001){Ouchi}, {Shimasaku}, {Okamura}, {Doi},
  {Furusawa}, {Hamabe}, {Kimura}, {Komiyama}, {Miyazaki}, {Miyazaki}, {Nakata},
  {Sekiguchi}, {Yagi}, \& {Yasuda}}]{ouchietal2001}
{Ouchi} M., {Shimasaku} K., {Okamura} S., {Doi} M., {Furusawa} H., {Hamabe} M.,
  {Kimura} M., {Komiyama} Y., {Miyazaki} M., {Miyazaki} S., {Nakata} F.,
  {Sekiguchi} M., {Yagi} M., {Yasuda} N., 2001, \apjl, 558, L83

\bibitem[{{Overzier} {et~al.}(2008){Overzier}, {Heckman}, {Kauffmann},
  {Seibert}, {Rich}, {Basu-Zych}, {Lotz}, {Aloisi}, {Charlot}, {Hoopes},
  {Martin}, {Schiminovich}, \& {Madore}}]{overzieretal2008}
{Overzier} R.~A., {Heckman} T.~M., {Kauffmann} G., {Seibert} M., {Rich} R.~M.,
  {Basu-Zych} A., {Lotz} J., {Aloisi} A., {Charlot} S., {Hoopes} C., {Martin}
  D.~C., {Schiminovich} D., {Madore} B., 2008, \apj, 677, 37

\bibitem[{{Papaderos} {et~al.}(2008){Papaderos}, {Guseva}, {Izotov}, \&
  {Fricke}}]{papaderosetal2008}
{Papaderos} P., {Guseva} N.~G., {Izotov} Y.~I., {Fricke} K.~J., 2008, \aap,
  491, 113

\bibitem[{{Papaderos} {et~al.}(2002){Papaderos}, {Izotov}, {Thuan}, {Noeske},
  {Fricke}, {Guseva}, \& {Green}}]{papaderosetal2002}
{Papaderos} P., {Izotov} Y.~I., {Thuan} T.~X., {Noeske} K.~G., {Fricke} K.~J.,
  {Guseva} N.~G., {Green} R.~F., 2002, \aap, 393, 461

\bibitem[{{Papaderos} {et~al.}(1996){Papaderos}, {Loose}, {Fricke}, \&
  {Thuan}}]{papaderosetal1996}
{Papaderos} P., {Loose} H.-H., {Fricke} K.~J., {Thuan} T.~X., 1996, \aap, 314,
  59

\bibitem[{{Pascarelle} {et~al.}(1998){Pascarelle}, {Lanzetta}, \&
  {Fern{\'a}ndez-Soto}}]{pascarelleetal1998}
{Pascarelle} S.~M., {Lanzetta} K.~M., {Fern{\'a}ndez-Soto} A., 1998, \apjl,
  508, L1

\bibitem[{{Pentericci} {et~al.}(2009){Pentericci}, {Grazian}, {Fontana},
  {Castellano}, {Giallongo}, {Salimbeni}, \& {Santini}}]{pentericcietal2009}
{Pentericci} L., {Grazian} A., {Fontana} A., {Castellano} M., {Giallongo} E.,
  {Salimbeni} S., {Santini} P., 2009, \aap, 494, 553

\bibitem[{{Pettini} {et~al.}(2001){Pettini}, {Shapley}, {Steidel}, {Cuby},
  {Dickinson}, {Moorwood}, {Adelberger}, \& {Giavalisco}}]{pettinietal2001}
{Pettini} M., {Shapley} A.~E., {Steidel} C.~C., {Cuby} J.-G., {Dickinson} M.,
  {Moorwood} A.~F.~M., {Adelberger} K.~L., {Giavalisco} M., 2001, \apj, 554,
  981

\bibitem[{{Phillips} {et~al.}(1997){Phillips}, {Guzman}, {Gallego}, {Koo},
  {Lowenthal}, {Vogt}, {Faber}, \& {Illingworth}}]{phillipsetal1997}
{Phillips} A.~C., {Guzman} R., {Gallego} J., {Koo} D.~C., {Lowenthal} J.~D.,
  {Vogt} N.~P., {Faber} S.~M., {Illingworth} G.~D., 1997, \apj, 489, 543

\bibitem[{{Ravindranath} {et~al.}(2006){Ravindranath}, {Giavalisco},
  {Ferguson}, {Conselice}, {Katz}, {Weinberg}, {Lotz}, {Dickinson}, {Fall},
  {Mobasher}, \& {Papovich}}]{ravindranathetal2006}
{Ravindranath} S., {Giavalisco} M., {Ferguson} H.~C., {Conselice} C., {Katz}
  N., {Weinberg} M., {Lotz} J., {Dickinson} M., {Fall} S.~M., {Mobasher} B.,
  {Papovich} C., 2006, \apj, 652, 963

\bibitem[{{Reddy} \& {Steidel}(2009)}]{reddyetal2009}
{Reddy} N.~A., {Steidel} C.~C., 2009, \apj, 692, 778

\bibitem[{{Richards} {et~al.}(2005){Richards}, {Croom}, {Anderson},
  {Bland-Hawthorn}, {Boyle}, {De Propris}, {Drinkwater}, {Fan}, {Gunn},
  {Ivezi{\'c}}, {Jester}, {Loveday}, {Meiksin}, {Miller}, \&
  et~al.}]{richardsetal2005}
{Richards} G.~T., {Croom} S.~M., {Anderson} S.~F., {Bland-Hawthorn} J., {Boyle}
  B.~J., {De Propris} R., {Drinkwater} M.~J., {Fan} X., {Gunn} J.~E.,
  {Ivezi{\'c}} {\v Z}., {Jester} S., {Loveday} J., {Meiksin} A., {Miller} L.,
  et~al., 2005, \mnras, 360, 839

\bibitem[{{Ryan} {et~al.}(2007){Ryan}, {De Robertis}, {Virani}, {Laor}, \&
  {Dawson}}]{ryanetal2007}
{Ryan} C.~J., {De Robertis} M.~M., {Virani} S., {Laor} A., {Dawson} P.~C.,
  2007, \apj, 654, 799

\bibitem[{{Sarzi} {et~al.}(2006){Sarzi}, {Falc{\'o}n-Barroso}, {Davies},
  {Bacon}, {Bureau}, {Cappellari}, {de Zeeuw}, {Emsellem}, {Fathi},
  {Krajnovi{\'c}}, {Kuntschner}, {McDermid}, \& {Peletier}}]{sarzietal2006}
{Sarzi} M., {Falc{\'o}n-Barroso} J., {Davies} R.~L., {Bacon} R., {Bureau} M.,
  {Cappellari} M., {de Zeeuw} P.~T., {Emsellem} E., {Fathi} K., {Krajnovi{\'c}}
  D., {Kuntschner} H., {McDermid} R.~M., {Peletier} R.~F., 2006, \mnras, 366,
  1151

\bibitem[{{Schawinski} {et~al.}(2009){Schawinski}, {Lintott}, {Thomas},
  {Sarzi}, {Andreescu}, {Bamford}, {Kaviraj}, {Khochfar}, {Land}, {Murray},
  {Nichol}, {Raddick}, {Slosar}, {Szalay}, {Vandenberg}, \&
  {Yi}}]{schawinskietal2009}
{Schawinski} K., {Lintott} C., {Thomas} D., {Sarzi} M., {Andreescu} D.,
  {Bamford} S.~P., {Kaviraj} S., {Khochfar} S., {Land} K., {Murray} P.,
  {Nichol} R.~C., {Raddick} M.~J., {Slosar} A., {Szalay} A., {Vandenberg} J.,
  {Yi} S.~K., 2009, \mnras, 690

\bibitem[{{Schawinski} {et~al.}(2007){Schawinski}, {Thomas}, {Sarzi},
  {Maraston}, {Kaviraj}, {Joo}, {Yi}, \& {Silk}}]{schawinskietal2007}
{Schawinski} K., {Thomas} D., {Sarzi} M., {Maraston} C., {Kaviraj} S., {Joo}
  S.-J., {Yi} S.~K., {Silk} J., 2007, \mnras, 382, 1415

\bibitem[{{Schmitt}(2006)}]{schmitt2006}
{Schmitt} H., 2006, in HST Proposal, pp. 10880--+

\bibitem[{{Shapley} {et~al.}(2001){Shapley}, {Steidel}, {Adelberger},
  {Dickinson}, {Giavalisco}, \& {Pettini}}]{shapleyetal2001}
{Shapley} A.~E., {Steidel} C.~C., {Adelberger} K.~L., {Dickinson} M.,
  {Giavalisco} M., {Pettini} M., 2001, \apj, 562, 95

\bibitem[{{Shapley} {et~al.}(2003){Shapley}, {Steidel}, {Pettini}, \&
  {Adelberger}}]{shapleyetal2003}
{Shapley} A.~E., {Steidel} C.~C., {Pettini} M., {Adelberger} K.~L., 2003, \apj,
  588, 65

\bibitem[{{Skibba} {et~al.}(2008){Skibba}, {Bamford}, {Nichol}, {Lintott},
  {Andreescu}, {Edmondson}, {Murray}, {Raddick}, {Schawinski}, {Slosar},
  {Szalay}, {Thomas}, \& {Vandenberg}}]{skibbaetal2008}
{Skibba} R.~A., {Bamford} S.~P., {Nichol} R.~C., {Lintott} C.~J., {Andreescu}
  D., {Edmondson} E.~M., {Murray} P., {Raddick} M.~J., {Schawinski} K.,
  {Slosar} A., {Szalay} A.~S., {Thomas} D., {Vandenberg} J., 2008,
  ArXiv:0811.3970

\bibitem[{{Slosar} {et~al.}(2009){Slosar}, {Land}, {Bamford}, {Lintott},
  {Andreescu}, {Murray}, {Nichol}, {Raddick}, {Schawinski}, {Szalay}, {Thomas},
  \& {Vandenberg}}]{slosaretal2009}
{Slosar} A., {Land} K., {Bamford} S., {Lintott} C., {Andreescu} D., {Murray}
  P., {Nichol} R., {Raddick} M.~J., {Schawinski} K., {Szalay} A., {Thomas} D.,
  {Vandenberg} J., 2009, \mnras, 392, 1225

\bibitem[{{Spergel} {et~al.}(2007){Spergel}, {Bean}, {Dor{\'e}}, \&
  {Nolta}}]{spergeletal2007}
{Spergel} D.~N., {Bean} R., {Dor{\'e}} O., {Nolta} M.~R., 2007, \apjs, 170, 377

\bibitem[{{Stanway} {et~al.}(2003){Stanway}, {Bunker}, \&
  {McMahon}}]{stanwayetal2003}
{Stanway} E.~R., {Bunker} A.~J., {McMahon} R.~G., 2003, \mnras, 342, 439

\bibitem[{{Steidel} {et~al.}(2000){Steidel}, {Adelberger}, {Shapley},
  {Pettini}, {Dickinson}, \& {Giavalisco}}]{steideletal2000}
{Steidel} C.~C., {Adelberger} K.~L., {Shapley} A.~E., {Pettini} M., {Dickinson}
  M., {Giavalisco} M., 2000, \apj, 532, 170

\bibitem[{{Steidel} {et~al.}(2003){Steidel}, {Adelberger}, {Shapley},
  {Pettini}, {Dickinson}, \& {Giavalisco}}]{steideletal2003}
---, 2003, \apj, 592, 728

\bibitem[{{Stoughton} {et~al.}(2002){Stoughton}, {Lupton}, {Bernardi},
  {Blanton}, {Burles}, {Castander}, {Connolly}, {York}, {Zehavi}, \&
  {Zheng}}]{stoughtonetal2002}
{Stoughton} C., {Lupton} R.~H., {Bernardi} M., {Blanton} M.~R., {Burles} S.,
  {Castander} F.~J., {Connolly} A.~J., {York} D.~G., {Zehavi} I., {Zheng} W.,
  2002, \aj, 123, 485

\bibitem[{{Strauss} {et~al.}(2002){Strauss}, {Weinberg}, \&
  {Lupton}}]{straussetal2002}
{Strauss} M.~A., {Weinberg} D.~H., {Lupton} R.~H., 2002, \aj, 124, 1810

\bibitem[{{Taniguchi}(2004)}]{taniguchi2004}
{Taniguchi} Y., 2004, in Studies of Galaxies in the Young Universe with New
  Generation Telescope, {Arimoto} N., {Duschl} W.~J., eds., pp. 107--111

\bibitem[{{Thomas} {et~al.}(2005){Thomas}, {Maraston}, {Bender}, \& {Mendes de
  Oliveira}}]{thomasetal2005}
{Thomas} D., {Maraston} C., {Bender} R., {Mendes de Oliveira} C., 2005, \apj,
  621, 673

\bibitem[{{Thommes} {et~al.}(1998){Thommes}, {Meisenheimer}, {Fockenbrock},
  {Hippelein}, {Roeser}, \& {Beckwith}}]{thommesetal1998}
{Thommes} E., {Meisenheimer} K., {Fockenbrock} R., {Hippelein} H., {Roeser}
  H.-J., {Beckwith} S., 1998, \mnras, 293, L6

\bibitem[{{Tremaine} {et~al.}(2002){Tremaine}, {Gebhardt}, {Bender}, {Bower},
  {Dressler}, {Faber}, {Filippenko}, {Green}, {Grillmair}, {Ho}, {Kormendy},
  {Lauer}, {Magorrian}, {Pinkney}, \& {Richstone}}]{tremaineetal2002}
{Tremaine} S., {Gebhardt} K., {Bender} R., {Bower} G., {Dressler} A., {Faber}
  S.~M., {Filippenko} A.~V., {Green} R., {Grillmair} C., {Ho} L.~C., {Kormendy}
  J., {Lauer} T.~R., {Magorrian} J., {Pinkney} J., {Richstone} D., 2002, \apj,
  574, 740

\bibitem[{{Tremonti} {et~al.}(2004){Tremonti}, {Heckman}, {Kauffmann},
  {Brinchmann}, {Charlot}, {White}, {Seibert}, {Peng}, {Schlegel}, {Uomoto},
  {Fukugita}, \& {Brinkmann}}]{tremontietal2004}
{Tremonti} C.~A., {Heckman} T.~M., {Kauffmann} G., {Brinchmann} J., {Charlot}
  S., {White} S.~D.~M., {Seibert} M., {Peng} E.~W., {Schlegel} D.~J., {Uomoto}
  A., {Fukugita} M., {Brinkmann} J., 2004, \apj, 613, 898

\bibitem[{{Vaduvescu} {et~al.}(2007){Vaduvescu}, {McCall}, \&
  {Richer}}]{vaduvescuetal2007}
{Vaduvescu} O., {McCall} M.~L., {Richer} M.~G., 2007, \aj, 134, 604

\bibitem[{{Veilleux} \& {Osterbrock}(1987)}]{veilleuxandosterbrock1987}
{Veilleux} S., {Osterbrock} D.~E., 1987, \apjs, 63, 295

\bibitem[{{Venemans} {et~al.}(2005){Venemans}, {R{\"o}ttgering}, {Miley},
  {Kurk}, {De Breuck}, {Overzier}, {van Breugel}, {Carilli}, {Ford}, {Heckman},
  {Pentericci}, \& {McCarthy}}]{venemansetal2005}
{Venemans} B.~P., {R{\"o}ttgering} H.~J.~A., {Miley} G.~K., {Kurk} J.~D., {De
  Breuck} C., {Overzier} R.~A., {van Breugel} W.~J.~M., {Carilli} C.~L., {Ford}
  H., {Heckman} T., {Pentericci} L., {McCarthy} P., 2005, \aap, 431, 793

\bibitem[{{Voges} {et~al.}(1999){Voges}, {Aschenbach}, {Boller},
  {Br{\"a}uninger}, {Tr{\"u}mper}, \& {Zimmermann}}]{vogesetal1999}
{Voges} W., {Aschenbach} B., {Boller} T., {Br{\"a}uninger} H., {Tr{\"u}mper}
  J., {Zimmermann} H.~U., 1999, \aap, 349, 389

\bibitem[{{Williams} {et~al.}(2002){Williams}, {Pogge}, \&
  {Mathur}}]{williamsetal2002}
{Williams} R.~J., {Pogge} R.~W., {Mathur} S., 2002, \aj, 124, 3042

\bibitem[{{Yabe} {et~al.}(2009){Yabe}, {Ohta}, {Iwata}, {Sawicki}, {Tamura},
  {Akiyama}, \& {Aoki}}]{yabeetal2008}
{Yabe} K., {Ohta} K., {Iwata} I., {Sawicki} M., {Tamura} N., {Akiyama} M.,
  {Aoki} K., 2009, \apj, 693, 507

\bibitem[{{Yasuda} {et~al.}(2001){Yasuda}, {Fukugita}, {Narayanan}, \&
  {Lupton}}]{yasudaetal2001}
{Yasuda} N., {Fukugita} M., {Narayanan} V.~K., {Lupton} R.~H., 2001, \aj, 122,
  1104

\bibitem[{{York} {et~al.}(2000){York}, {Adelman}, {Anderson}, {Anderson},
  {Annis}, {Bahcall}, \& {Bakken}}]{yorketal2000}
{York} D.~G., {Adelman} J., {Anderson} Jr. J.~E., {Anderson} S.~F., {Annis} J.,
  {Bahcall} N.~A., {Bakken} J.~A., 2000, \aj, 120, 1579

\bibitem[{{Zhou} {et~al.}(2006){Zhou}, {Wang}, {Yuan}, {Lu}, {Dong}, {Wang}, \&
  {Lu}}]{zhouetal2006}
{Zhou} H., {Wang} T., {Yuan} W., {Lu} H., {Dong} X., {Wang} J., {Lu} Y., 2006,
  \apjs, 166, 128

\bibitem[{{Zwicky}(1965)}]{zwicky1965}
{Zwicky} F., 1965, \apj, 142, 1293

\end{thebibliography}
\end{document}